\documentclass[prc,aps,twocolumn,showpacs,amssymb,superscriptaddress,fleqn]{revtex4-1}

\usepackage{amsmath}
\usepackage{color}
\usepackage{graphicx}
\usepackage{dcolumn}
\usepackage{mathtools}
\usepackage{relsize}
\usepackage{float}
\usepackage{braket}

\DeclareSymbolFont{largesymbol}{OMX}{yhex}{m}{n}
\DeclareMathAccent{\Widehat}{\mathord}{largesymbol}{"62}

\begin{document}

\title{The contribution of chiral three-body forces \\
to the monopole component of the effective shell-model Hamiltonian}

\author{Y. Z. Ma}
\affiliation{School of Physics, and State Key Laboratory of Nuclear
  Physics and Technology, \\
Peking University, Beijing 100871, China}
\author{L. Coraggio}
\affiliation{Istituto Nazionale di Fisica Nucleare, \\
Complesso Universitario di Monte  S. Angelo, Via Cintia, I-80126 Napoli, Italy}
\author{L. De Angelis}
\affiliation{Istituto Nazionale di Fisica Nucleare, \\
Complesso Universitario di Monte  S. Angelo, Via Cintia, I-80126 Napoli, Italy}
\author{T. Fukui}
\affiliation{Istituto Nazionale di Fisica Nucleare, \\
Complesso Universitario di Monte  S. Angelo, Via Cintia, I-80126 Napoli, Italy}
\author{A. Gargano}
\affiliation{Istituto Nazionale di Fisica Nucleare, \\
Complesso Universitario di Monte  S. Angelo, Via Cintia, I-80126 Napoli, Italy}
\author{N. Itaco}
\affiliation{Istituto Nazionale di Fisica Nucleare, \\ 
Complesso Universitario di Monte  S. Angelo, Via Cintia, I-80126 Napoli, Italy}
\affiliation{Dipartimento di Matematica e Fisica, Universit\`a degli
  Studi della Campania ``Luigi Vanvitelli'', viale Abramo Lincoln 5 -
  I-81100 Caserta, Italy}
\author{F. R. Xu}
\affiliation{School of Physics, and State Key Laboratory of Nuclear
  Physics and Technology, \\
Peking University, Beijing 100871, China}
 
\begin{abstract}
We present a study of the role played by realistic three-body forces
in providing a reliable monopole component of the effective
shell-model Hamiltonian.
To this end, starting from a nuclear potential built up within the
chiral perturbation theory, we derive effective shell-model
Hamiltonians with and without the contribution of the three-body
potential and compare the results of shell-model calculations with a
set of observables that evidence shell-evolution properties.
The testing ground of our investigation are nuclei belonging to $fp$
shell, since the shell-evolution towards shell closures in $^{48}$Ca
and $^{56}$Ni provides a paradigm for shell-model Hamiltonians.
Our analysis shows that only by including contributions of the three-body
force the monopole component of the effective shell-model Hamiltonian
is then able to reproduce the experimental shell evolution towards and
beyond the closure at $N=28$.
\end{abstract}


\maketitle

\section{Introduction}
\label{intro}
The evolution of the nuclear spectroscopic properties along isotopic
and isotonic chains, towards the formation of magic numbers, is the
feature that reveals the central role of the nuclear shell model (SM)
and its success during the past 70 years
\cite{Mayer49,Haxel49,Mayer55}. 
Consequently, it is very desirable that effective Hamiltonians,
which are employed to study the nuclear structure in the framework of
the shell model, should be able to reproduce the observed shell
evolution and closures.

Zuker and coworkers have extensively investigated the
properties of the two-body matrix elements (TBMEs) of the residual
interaction derived from realistic potentials by way of many-body
perturbation theory \cite{Kuo66}, and, having performed a multipole
decomposition of realistic SM Hamiltonians, have shown that
their monopole component needs to be modified in order to reproduce
the evolution of shell closures as a function of the number of valence
nucleons \cite{Caurier94,Duflo99,Zuker00}. 
They have inferred that this should trace back to the
lack of a three-nucleon force (3NF) in the nuclear realistic
potentials employed to derive the effective SM Hamiltonian $H_{\rm
  eff}$, affecting its monopole component that, consequently, has to
be corrected \cite{Zuker03}.

Extensive direct investigations about the role of 3NFs in realistic
$H_{\rm eff}$s have been carried out by Schwenk and coworkers, who have
performed studies of oxygen
\cite{Otsuka10,Holt13c,Hebeler15,Simonis16,Simonis17} and calcium
\cite{Holt12,Holt14,Hebeler15,Simonis17} isotopic chains.
In the aforementioned works, the $H_{\rm eff}$s blue have been derived
starting from nuclear potentials built up within the chiral
perturbative expansion and softened by way of the $V_{\rm low-k}$
technique \cite{Bogner02,Bogner03} or the similarity
renormalization-group (SRG) approach \cite{Bogner07,Bogner10}, and the
results have supported the need of introducing three-body forces to
reproduce the experimental behavior of the ground-state and yrast
excitation energies as a function of the valence-nucleon number.

In order to investigate the role played by three-body forces in
driving the shell evolution, we have found inspiration from the
calculation of the effective single-particle energies (ESPEs) for
$p$-shell nuclei, whose results we have presented in
Ref. \cite{Coraggio18a}.
More precisely, we have found that the ESPEs calculated from the $H_{\rm
  eff}$ that includes contributions from both two- and three-body
chiral potentials, provide a constant energy-splitting of the
spin-orbit partners $0p_{3/2},0p_{1/2}$ as a function of the mass
number $A$.
This splitting characterizes the correct reproduction of the subshell
closure at $Z,N=6$ observed in $^{12}$C, at variance with the result
we have obtained omitting the contribution of the 3NF.
As a matter of fact, the relative ESPE rapidly drops down if only the
two-nucleon force (2NF) is included, and becomes even negative around
$A=8$.
Then, the reproduction of the shell closure deteriorates, namely the
observed energy of the $^{12}$C yrast $J^{\pi}=2^+$ state is
underestimated by $\sim 1$ MeV.

Since the ESPE  of a level is calculated in terms of the bare
single-particle (SP) energy and the monopole part of the TBMEs
\cite{Utsuno99}, it is clear that the above mentioned results point to
an intimate relationship between 3NF and the monopole component of
$H_{\rm eff}$.

On the above grounds, we devote the present paper to studying this
connection choosing, as a testing ground, the nuclei belonging to the
$fp$ shell, namely those that can be described in terms
  of the degrees of freedom of valence nucleons outside doubly-closed
$^{40}$Ca, interacting in the model space composed by $0f1p$
orbitals.
This region represents a paradigm to investigate the shell evolution
within the shell model, since, as is well-known, the spin-orbit
component of the SM mean field separates the $0f_{7/2}$ orbital from
the others leading to the appearance of the magic number
  $Z,N=28$ and, consequently, of the two doubly-magic nuclei $^{48}$Ca
  and $^{56}$Ni.

The starting point of our calculation is a nuclear potential based on
chiral perturbation theory (ChPT) \cite{Epelbaum05,Machleidt11}, a
choice that is motivated by two main considerations.
\begin{enumerate}
\item[a)] First, within this class of potentials long-range forces are
  ruled by the symmetries of low-energy quantum chromodynamics (QCD) -
  in particular the spontaneously broken chiral symmetry - and the
  short-range dynamics is absorbed into a complete basis of contact
  terms that are proportional to low-energy constants (LECs) fitted to
  two-nucleon data.
\item[b)] The second major characteristic of ChPT is that nuclear 2NF
  and many-body forces are generated on an equal footing
  \cite{Weinberg92,vanKolck94,Machleidt11}, since most interaction
  vertices that appear in the 3NF and in the four-nucleon force (4NF)
  also occur in the 2NF.
\end{enumerate}
For the sake of completeness, we point out that, as in
Ref. \cite{Coraggio18a}, a high-precision 2NF potential derived within
the ChPT at next-to-next-to-next-to-leading order (N$^3$LO)
\cite{Entem02,Machleidt11} is considered in our calculation, without
any renormalization of its high-momentum components, juxtaposed with
a N$^2$LO 3NF potential, since this many-body contribution appears
from this order on.
Nowadays, these potentials are widely employed in nuclear theory
aiming to link the fundamental theory of strong interactions, the QCD,
to nuclear many-body phenomena.

Then, the $H_{\rm eff}$s for systems with one- and two-valence
nucleons outside the $^{40}$Ca core are derived by way of the
energy-independent linked-diagram perturbation theory \cite{Kuo90},
where 2NF-vertices diagrams are included up to third order and
contributions of 3NF up to first order in the perturbative expansion.

For those nuclei with a number of valence nucleons larger than 2 - we
will report calculations for $Z=20$, 22, 24, 26, and 28 up to $N=40$ -
the effect of many-body correlations is taken into account by
including the contributions of three-body diagrams calculated at
second order in perturbation theory \cite{Polls83}.
These correlations arise from the interaction via the two-body force
of the valence nucleons with excitations outside the model space
\cite{Ellis77}.
Since our SM code cannot manage three-body Hamiltonians, we have
derived a density-dependent two-body contribution at one-loop order
from the three-body correlation diagrams, summing over the
partially-filled model-space orbitals.

A description of the perturbative approach to the derivation of
our effective SM Hamiltonian is reported in Section \ref{outline},
where the perturbative properties are also discussed in some detail.
In Section \ref{results} we introduce first the results of the
calculation of the ESPEs, in order to analyze the properties of the
monopole component of the effective Hamiltonians, obtained with and
without the contribution from a chiral 3NF.
Then, we compare the results of the full diagonalization of these
$H_{\rm eff}$s with observables that are sensitive to the shell
evolution of $fp$ isotopic chains.
We focus on the evolution of collectivity in $N=28$ isotones too, that
is a key point to evaluate the balance between the monopole and
quadrupole components of the effective SM Hamiltonian.
Finally, in Section \ref{conclusions} we draw the conclusions of our
study and the outlook of our future work.

\section{Outline of calculations}
\label{outline}
As mentioned in the Introduction, we choose, as 2NF, the chiral
N$^3$LO potential derived by Entem and Machleidt in
Ref. \cite{Entem02}, and as 3NF a chiral N$^2$LO potential, which
shares the regulator function of a nonlocal form and some of the LECs
with the 2NF.
It is worth pointing out that the N$^2$LO 3NF is composed of
three components, namely the two-pion ($2\pi$) exchange term
$V_{3NF}^{(2\pi)}$, the one-pion ($1\pi$) exchange plus contact term
$V_{3NF}^{(1\pi)}$, and the contact term $V_{3NF}^{\textrm{(ct)}}$.

For the sake of consistency, the $c_1$, $c_3$, and $c_4$ LECs
appearing in $V_{\rm 3NF}^{(2\pi)}$, are the same as those in the
N$^3$LO 2NF, their values being determined by the
renormalization procedure that fits the nucleon-nucleon ($NN$) data
\cite{Machleidt11}.

Moreover, the 3NF 1$\pi$-exchange and contact terms are characterized
by two extra LECs (known as $c_D$ and $c_E$, respectively), which
cannot be constrained by two-body observables, but need to be
determined by reproducing observables in systems with mass $A>2$.

We adopt the same $c_D,c_E$ values employed in
Ref. \cite{Coraggio18a}, namely $c_D=-1$ and $c_E=-0.34$, that have
been determined by way of no-core shell model (NCSM) calculations
\cite{Navratil07a,Maris13}.
More precisely, in Ref. \cite{Navratil07a} it has been identified a
set of observables in light $p$-shell nuclei that are strongly sensitive to
the $c_D$ value in order to fix it, then $c_E$ has been constrained to
reproduce the binding energies of the $A=3$ system.

Details about the calculation of our 3NF matrix elements in the
harmonic-oscillator (HO) basis are reported in Appendix of
Ref. \cite{Coraggio18a}.
Note that the Coulomb potential is explicitly taken into account in
our calculations.

In the same paper, a comprehensive description of the derivation of
our effective SM Hamiltonians for one- and two-valence nucleon
systems, starting from 2NF and 3NF, can also be found, while in the
following we present only a short summary.

As mentioned before, our $H_{\rm eff}$ is derived in the model space
spanned by the four $0f1p$ proton and neutron orbitals outside
doubly-closed $^{40}$Ca.

To this end, an auxiliary one-body potential $U$ is introduced in
order to break up the Hamiltonian $H$ for a system of $A$ nucleons as
the sum of a one-body term $H_0$, which describes the independent
motion of the nucleons, and a residual interaction $H_1$:
\begin{eqnarray}
\label{smham}
 H&=&\sum_{i=1}^{A} \frac{p_i^2}{2m} + \sum_{i<j=1}^{A} V_{ij}^{\rm 2NF}
  + \sum_{i<j<k=1}^{A} V_{ijk}^{\rm 3NF} = \\
~ &= &  T + V^{\rm 2NF}+ V^{\rm 3NF}=
       (T+U)+ \nonumber \\ 
  ~&~& +(V^{\rm 2NF}-U)+ V^{\rm 3NF}= H_{0}+H^{\rm 2NF}_{1}+H^{\rm
       3NF}_{1}~~. \nonumber
\end{eqnarray}

\noindent
In our calculation we use the HO potential, $U=\frac{1}{2} m \omega ^2
r^2$, with an oscillator parameter $\hbar \omega = 11$ MeV, according
to the expression \cite{Blomqvist68} $\hbar \omega= 45 A^{-1/3} -25
A^{-2/3}$  for $A=40$.

Once $H_0$ has been introduced, the reduced model space is defined in
terms of a finite subset of $H_0$'s eigenvectors. 
The diagonalization of the many-body Hamiltonian in Eq. (\ref{smham})
within the infinite Hilbert space, that it is obviously unfeasible, is then
reduced to the solution of an eigenvalue problem for an effective 
Hamiltonian $H_{\rm eff}$ in a finite space.

Our approach to the derivation of $H_{\rm eff}$ is the time-dependent
perturbation theory \cite{Kuo90,Hjorth95,Coraggio12a}.
Namely, $H_{\rm eff}$ is expressed through the Kuo-Lee-Ratcliff (KLR)
folded-diagram expansion in terms of the vertex function
$\hat{Q}$-box, which is composed of irreducible valence-linked
diagrams \cite{Kuo71,Kuo81}.
We include in the $\hat{Q}$-box one- and two-body Goldstone diagrams
through third order in $H_1^{\rm 2NF}$ and up to first order in
$H_1^{\rm 3NF}$.

\begin{figure}[h]
\begin{center}
\includegraphics[scale=1.0,angle=0]{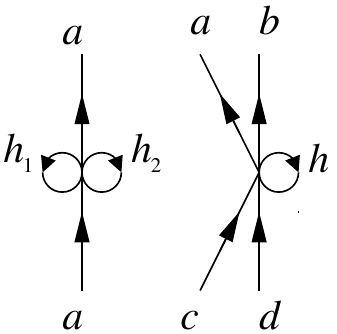}
\caption{First-order one- and two-body diagrams with a
  three-body-force vertex. See text for details.}
\label{1b2b3bf}
\end{center}
\end{figure}

In Fig. \ref{1b2b3bf} we report the contribution at first order in
perturbation theory to the single-particle component of the
$\hat{Q}$-box of a three-body potential, whose explicit expression is:

\begin{align}
\label{1b3bfeq}
&\langle j_a| 1{\rm b}_{3N} | j_a \rangle =  ~\nonumber \\
&\mathlarger{\sum}_{\substack{h_1,h_2{}\\J_{12}J}} ~
\frac{\hat{J}^2}{2 \hat{j_a}^2} \langle \left[
  (j_{h_1} j_{h_2})_{J_{12}},j_a \right]_{J}| V_{3N} | \left[
  (j_{h_1} j_{h_2})_{J_{12}},j_a \right]_{J} \rangle~~.
\end{align}

\noindent
The expression of the first-order two-body diagram with a $3N$ vertex,
shown in Fig. \ref{1b2b3bf}, is the following:

\begin{align}
\label{2b3bfeq}
&\langle (j_a j_b)_J| 2{\rm b}_{3N} | (j_c j_d)_J \rangle =  ~\nonumber \\
&\mathlarger{\sum}_{h,J'} ~
\frac{\hat{J'}^2}{\hat{J}^2} \langle \left[
  (j_{a} j_{b})_{J},j_h \right]_{J'}| V_{3N} | \left[
  (j_{c} j_{d})_{J},j_h \right]_{J'} \rangle~~,
\end{align}

The three-body matrix element (3BME) $\langle \left[ (j_{a}
  j_{b})_{J_{ab}},j_c \right]_{J}| V_{3N} | \left[ (j_{d}
  j_{e})_{J_{de}},j_f \right]_{J} \rangle$, expressed within the
proton-neutron formalism, is antisymmetrized but not normalized.

We recall that the expressions in Eqs. (\ref{1b3bfeq}) and
(\ref{2b3bfeq}) are the coefficients of the one-body and
two-body terms, respectively, arising from the normal-ordering
decomposition of the three-body component of a many-body
Hamiltonian \cite{HjorthJensen17}.

As mentioned in the Introduction, we include in the calculation of the
$\hat{Q}$-box also the effect of second-order three-body diagrams,
which, for those nuclei with more than 2 valence nucleons, account for
the interaction via the two-body force of the valence nucleons with
core excitations as well as with virtual intermediate nucleons
scattered above the model space. 

\begin{figure}[h]
\begin{center}
\includegraphics[scale=0.5,angle=0]{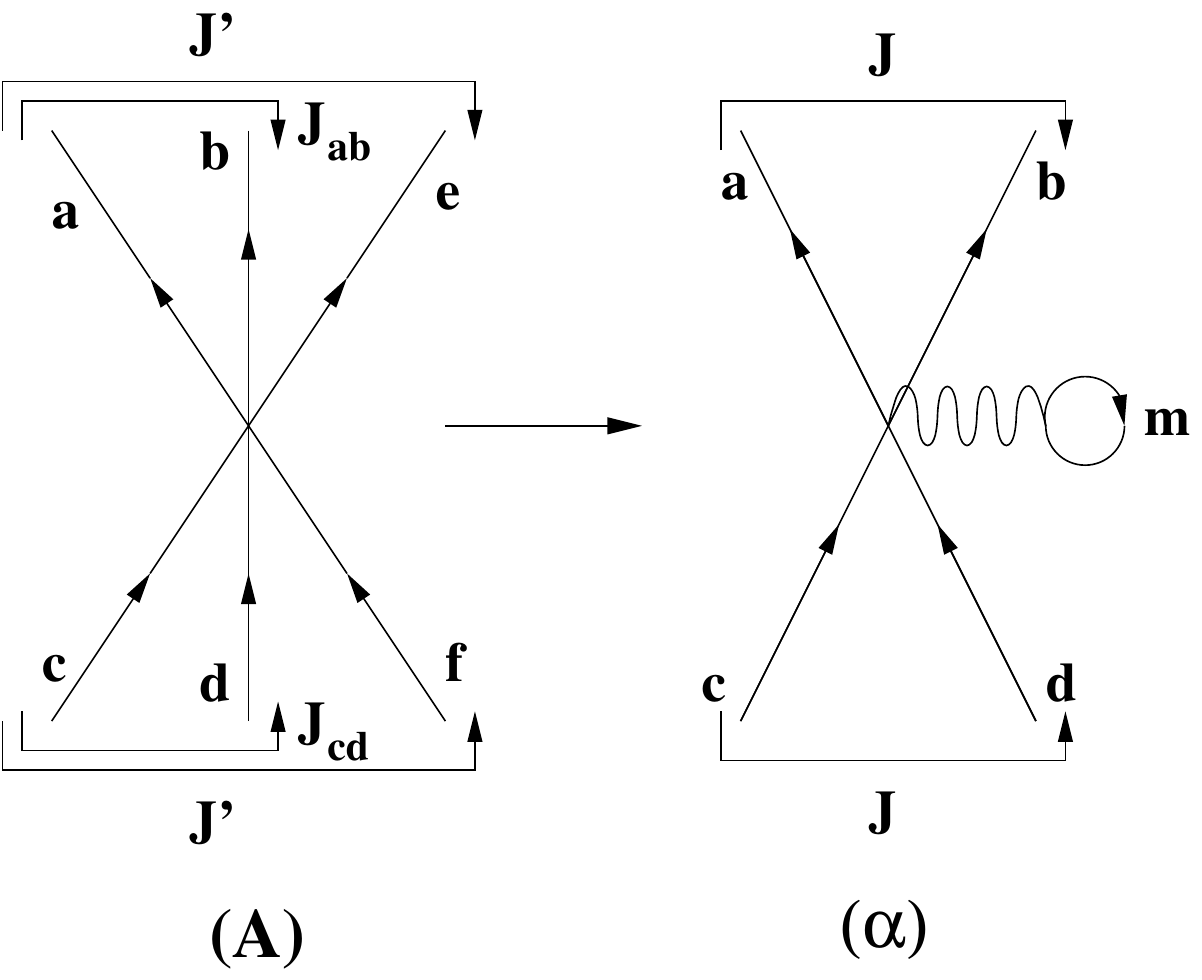}
\caption{Density-dependent two-body contribution that is
  obtained from a three-body one. $\alpha$ is obtained by
  summing over one incoming and outgoing particle of the three-body
  graph $A$ (see text for details).}
\label{3bf}
\end{center}
\end{figure}

 The SM code we employ \cite{ANTOINE} cannot perform the
diagonalization of a three-body $H_{\rm eff}$, so we derive from the
leading-order three-body contribution a density-dependent two-body term.
To this end, we calculate nine one-loop diagrams - the graph $(\alpha)$ in
Fig. \ref{3bf} - from the corresponding diagrams reported in Fig. 3 of
Ref. \cite{Polls83}.

Their explicit form, in terms of the three-body graph $(A)$, is the
same as in Eq. \ref{2b3bfeq}:

\begin{align}
\label{correq}
&\langle (j_a j_b)_J| V^{\alpha} | (j_c j_d)_J \rangle =  ~\nonumber \\
&\mathlarger{\sum}_{m,J'} ~ \rho_m
\frac{\hat{J'}^2}{\hat{J}^2} \langle \left[
  (j_{a} j_{b})_{J},j_m \right]_{J'} | V^A  | \left[
  (j_{c} j_{d})_{J},j_m \right]_{J'} \rangle~~,
\end{align}

\noindent
where the summation over $m$-index runs in the model space and the
expressions of the nine second-order diagrams $(A)$ are reported in
Appendix of Ref. \cite{Polls83}.
$\rho_m$ is the unperturbed occupation density of the orbital $j_m$
according to the number of valence nucleons.

Finally, the perturbative expression of the $\hat{Q}$-box contains one- and
two-body diagrams up to third order in the N$^3$LO 2NF
\cite{Coraggio12a}, one- and two-body first-order contributions in the
N$^2$LO 3NF \cite{Coraggio18a}, and a density-dependent two-body
contribution that accounts for three-body diagrams at second-order in
the N$^3$LO 2NF \cite{Ellis77,Polls83}.

It should be pointed out that the latter term will lead to the
derivation of specific effective shell-model Hamiltonians depending on
the number of valence protons and neutrons, that obviously differ only
for the two-body matrix elements.

The folded-diagram series is then summed up to all orders using the
Lee-Suzuki iteration method \cite{Suzuki80}.

We stress that the input chiral 2NF and 3NF have not been modified by
way of any renormalization procedure, and here we will show a few
details about the perturbative properties of the effective Hamiltonian.
A similar discussion about the perturbative expansion of the
$\hat{Q}$-box from N$^3$LO 2NF potential has been reported in
Ref. \cite{Coraggio12a}.

First, it should be pointed out that the truncation of the number of
intermediate states appearing in the perturbative expansion is the
same as in Ref. \cite{Coraggio12a}, i.e. the intermediate states whose
unperturbed excitation energy is greater than a fixed value
$E_{max}=N_{max} \hbar \omega$ are disregarded. 
As mentioned above, the value we have chosen for the HO parameter is
$\hbar \omega=11$ MeV. 
Because of our present limitation of the storage of the total number
of two-body matrix elements, we can include a maximum number of
intermediate states that do not exceed $N_{max}=18$.

\begin{figure}[h]
\begin{center}
\includegraphics[scale=0.30,angle=0]{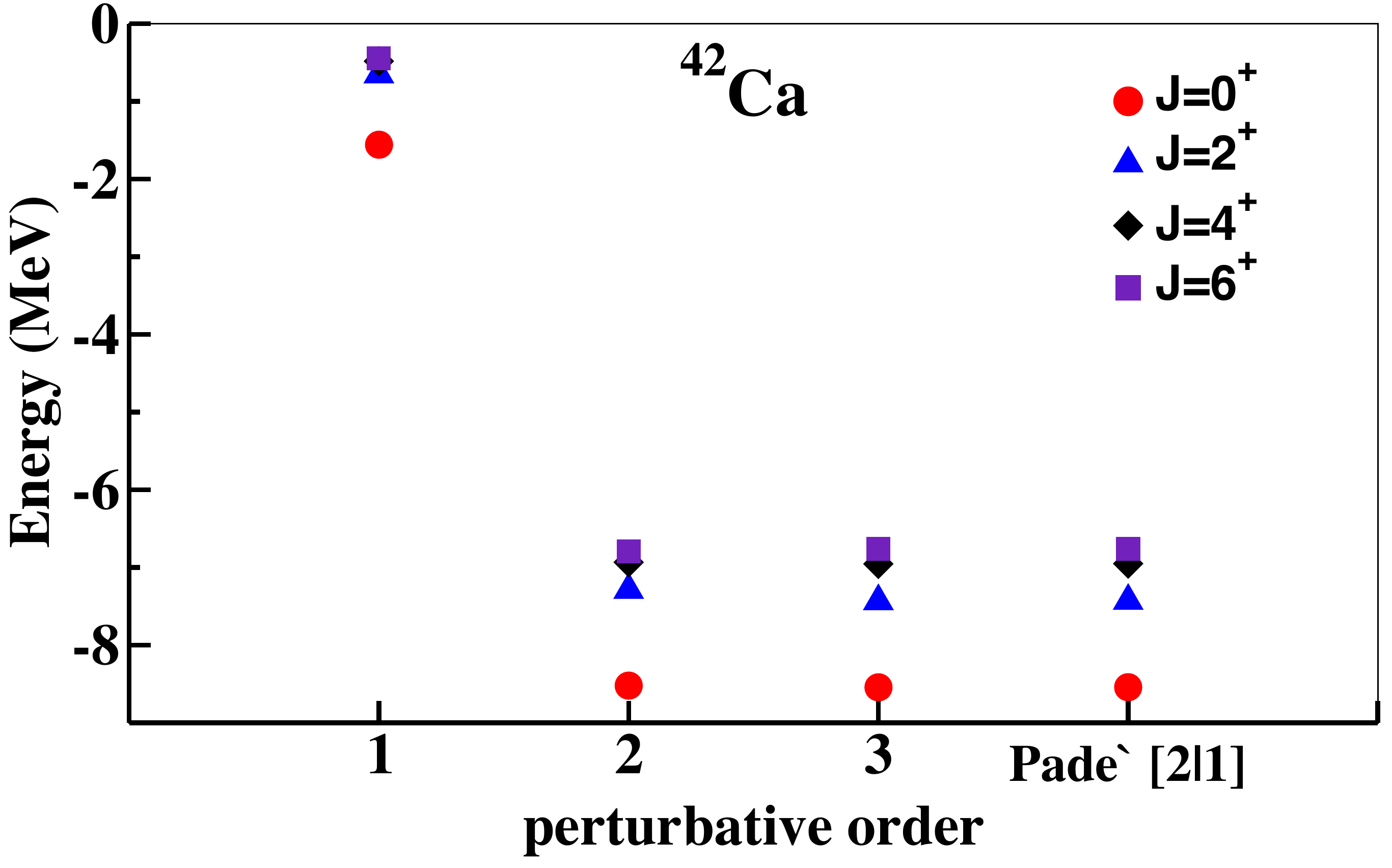}
\caption{Low-lying energy spectrum of $^{42}$Ca, obtained starting
  from $\hat{Q}$-boxes at first-, second-, and third-order in
  perturbation theory, and their Pad\'e approximant $[2|1]$. See text
  for details.}
\label{42Ca_obo}
\end{center}
\end{figure}

After these clarifying details, we present in Fig. \ref{42Ca_obo} the
first excited states of $^{42}$Ca spectrum, which have been obtained
employing $H_{\rm eff}$s with contributions of 3NF, and starting from
$\hat{Q}$-boxes at first-, second-, and third-order in perturbation
theory, and their Pad\'e approximant $[2|1]$\cite{Baker70}. 
We employ the Pad\'e approximant in order to obtain a better estimate
of the convergence value of the perturbation series
\cite{Coraggio12a}, as suggested in \cite{Hoffmann76}.
The number of intermediate states is the largest we can employ,
corresponding to $N_{max}=18$.

As can be seen, the results show a very satisfactory convergence of the
$H_{\rm eff}$ with respect to the order-by-order behavior of the
perturbative expansion.

We now move our focus to the issue of the dependence of $H_{\rm
  eff}$ with respect to the number of intermediate states included in the
calculation of second- and third-order diagrams.

\begin{figure}[h]
\begin{center}
\includegraphics[scale=0.30,angle=0]{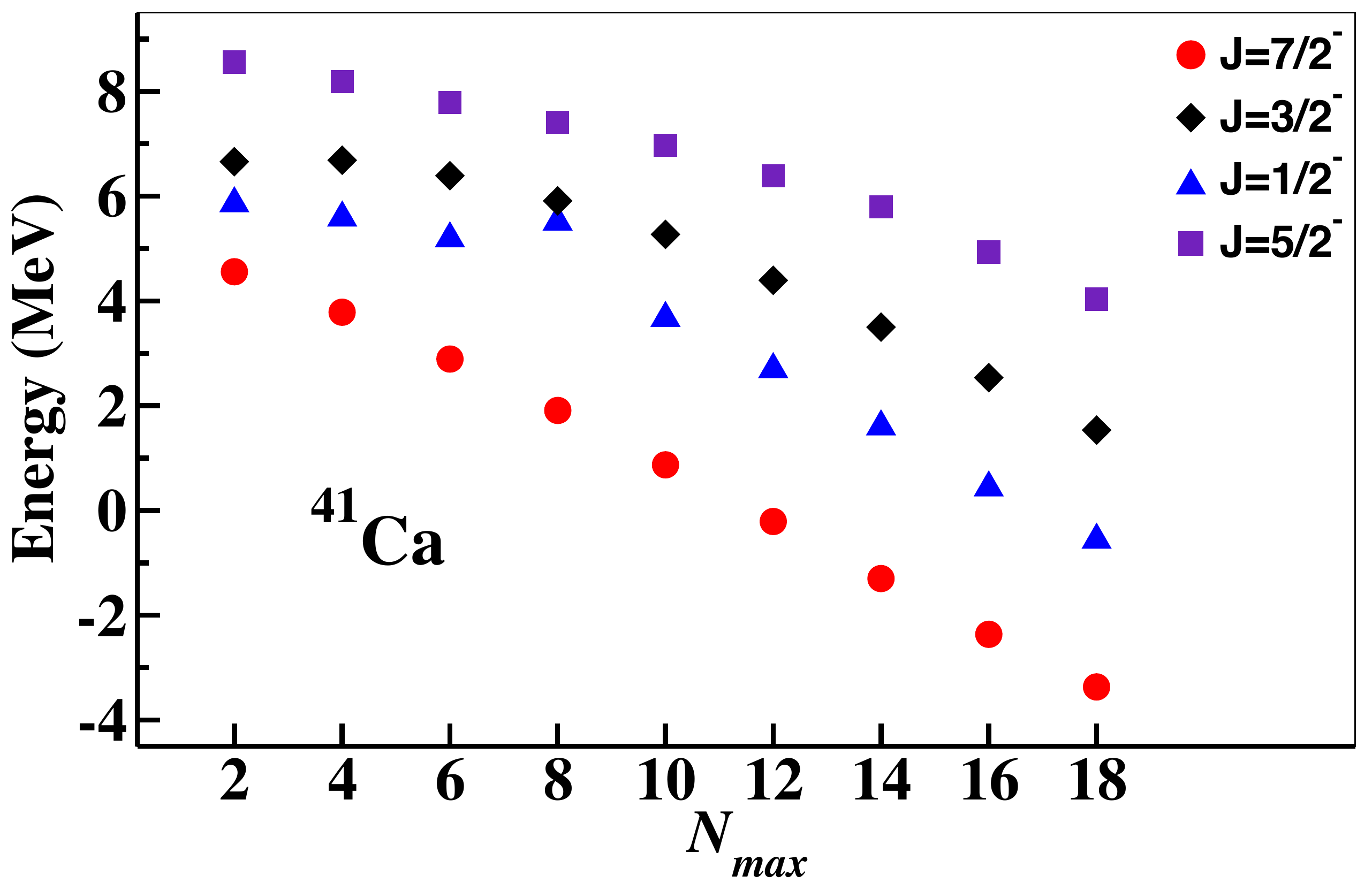}
\caption{ Low-lying energy spectrum of $^{41}$Ca relative to
  $^{40}$Ca as a function of $N_{\rm max}$ (see text for details).}
\label{41Ca_Nmax}
\end{center}
\end{figure}

In Fig. \ref{41Ca_Nmax} they are reported the energy spectra of
$^{41}$Ca, obtained from one-valence-neutron $H_{\rm
  eff}$s derived by employing the Pad\'e approximant $[2|1]$ of the
$\hat{Q}$-box, and including a number of intermediates states ranging
from $N_{max}=2$ to 18.

From the inspection of Fig. \ref{41Ca_Nmax}, it is evident that there
is no sign of convergence of the single-particle spectrum of
$^{41}$Ca up to $N_{max}=18$.
Since the cutoff of both 2NF and 3NF is slightly larger than 2.5 fm$^{-1}$
and we have chosen a value of the HO parameter to be equal to 11 MeV,
we estimate that we need at least $N_{max} \approx 24 - 26$ to reach the
convergence.
However, it can be clearly seen that from $N_{max} \approx 12 - 14$ on
the energy spacings are stable with respect to the increase in the
number of intermediate states.
This is an important feature, since the $H_{\rm eff}$ for one
valence-nucleon systems provides the SP energies for the SM
calculations, and it is highly desirable to obtain a convergent set of
theoretical SP energies to calculate excitation spectra of $fp$-shell
nuclei.

Actually, the fact that the SP energies which are calculated with
respect to the closed $^{40}$Ca do not converge with the increasing
number of intermediate states affects only the value of the
ground-state energy of open-shell systems.
Consequently, from now on we will employ, for our calculations,
SP spacings obtained from the theory while the value of the SP energy
of the $0f_{7/2}$ orbital is fixed at -1.1 MeV for protons and -8.4
MeV for neutrons, consistently with experimental values of $^{41}$Sc
and $^{41}$Ca \cite{Audi03}.

After the above considerations, we move to discuss the convergence of
two-valence-nucleon $H_{\rm eff}$ with respect to the number of
intermediate states.
As a matter of fact, this will be a test for our theoretical TBME,
since we have just observed that the SP energy spacings are convergent.

\begin{figure}[h]
\begin{center}
\includegraphics[scale=0.30,angle=0]{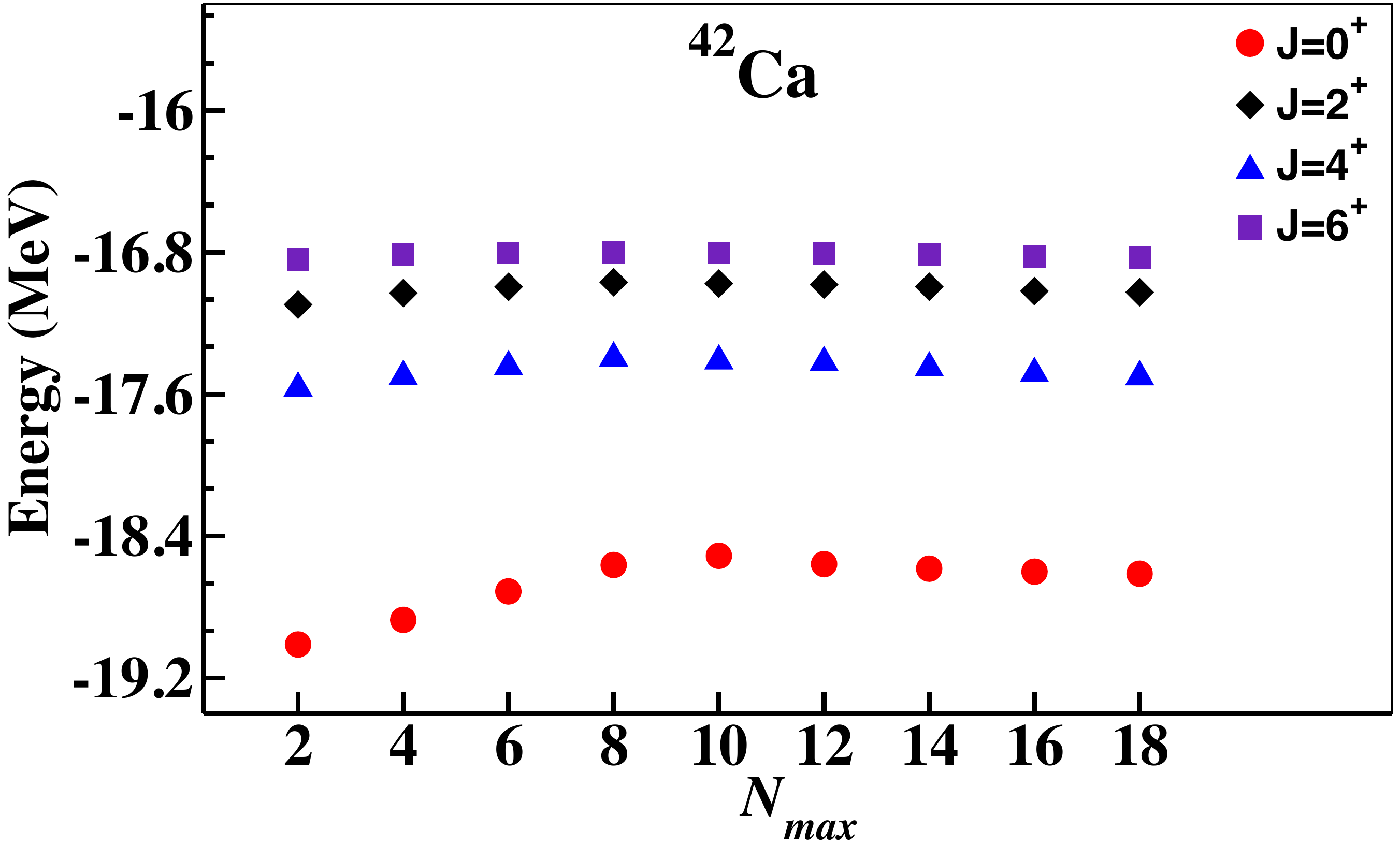}
\caption{Low-lying energy spectrum of $^{42}$Ca as a function of the
  number of intermediate states included in the perturbative
  calculation of the $\hat{Q}$-box. See text for details.}
\label{42Ca_Nmax}
\end{center}
\end{figure}

The calculated low-lying energy spectra of $^{42}$Ca, as a function of
$N_{max}$, are reported in Fig. \ref{42Ca_Nmax} up to $N_{max}=18$.
The Pad\`e approximant $[2|1]$ of the $\hat{Q}$-box has been
calculated to derive the $H_{\rm eff}$s, and the theoretical SP
spacings are considered relative to the experimental SP energy of the
$0f_{7/2}$ orbital, as mentioned before.

As it happens for $^{41}$Ca, we observe that also the $^{42}$Ca
spectrum converges from $N_{max}=12-14$ on.
This leads to the conclusion that both SP spacings and TBME of our
$H_{\rm eff}$, calculated with $N_{max}=18$, can be considered
substantially stable.

Besides the convergence behavior of our $H_{\rm eff}$, it is also
important to point out that, owing to the presence of the $-U$ term
in $H_1^{\rm 2NF}$, $U$-insertion diagrams arise in the
$\hat{Q}$-box, and that are responsible for controlling the $\hbar
\omega$ dependence introduced by the auxiliary potential $U$.

\begin{figure}[h]
\begin{center}
\includegraphics[scale=0.30,angle=0]{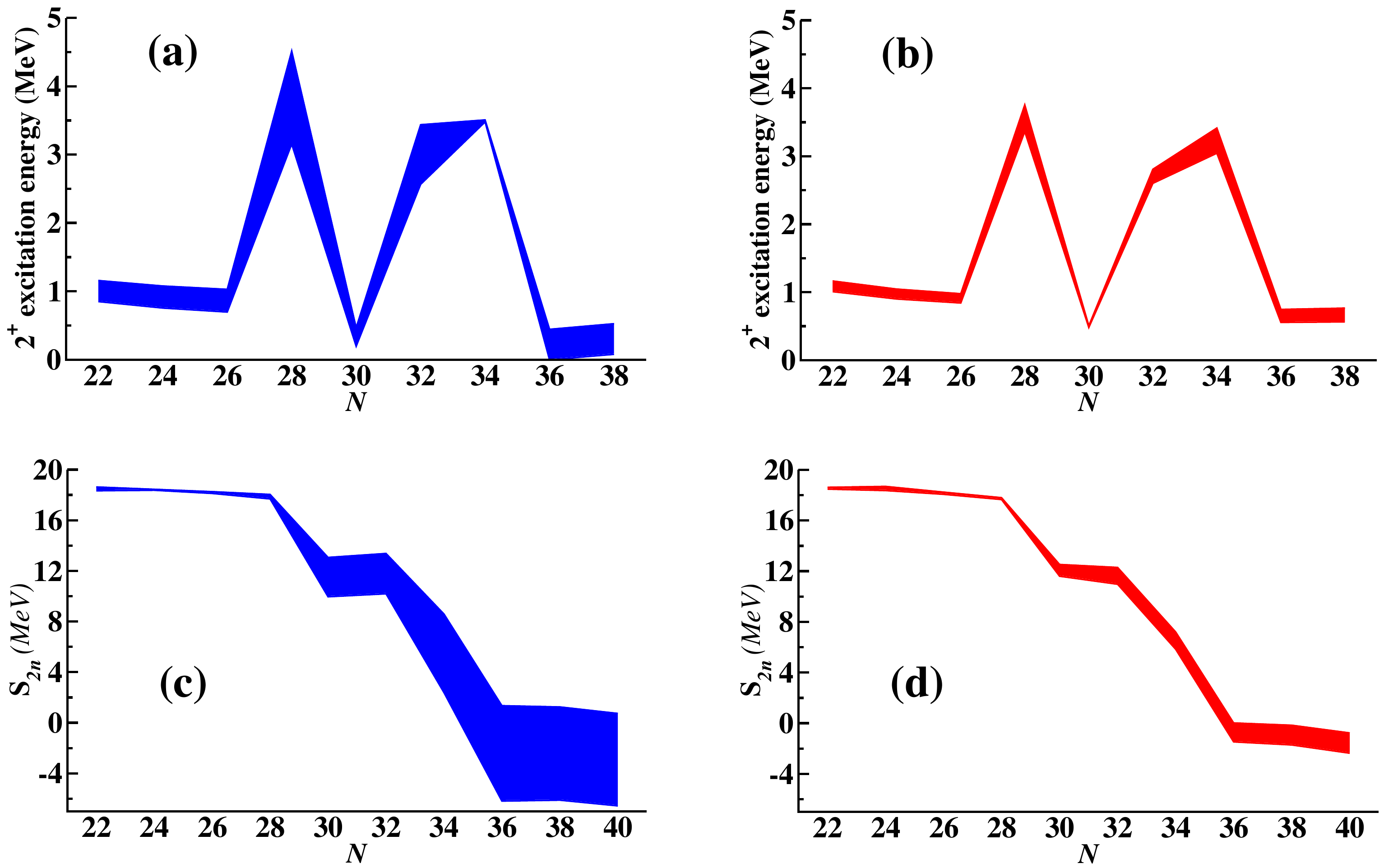}
\caption{Excitation energies of yrast $J^{\pi}=2^+$ states and
  $S_{2n}$ obtained for valuest of $\hbar \omega$ ranging from 10 to
  12 MeV. Blue bands corresponds to the $\hat{Q}$-boxes including
  $U$-insertion diagrams only at first order, red bands represent
  results obtained including $U$-insertion diagrams up to third-order
  in perturbation theory. See text for details.}
\label{hw_convergence}
\end{center}
\end{figure}

We have already addressed this issue in Ref. \cite{Coraggio12a} (see
Fig. 11 therein) and, in order to consider it within the present
study, we show the results of the calculated yrast $J^{\pi}=2^+$
excitation energies and two-neutron separation energies ($S_{2n}$) for
calcium isotopes up to $N=36$ in Fig. \ref{hw_convergence}, obtained
with different values of the HO parameter.

The results reported in Fig. \ref{hw_convergence} have been obtained
varying $\hbar \omega$ from 10 to 12 MeV.
The blue bands represent the variation that is obtained if only
first-order $U$-insertion diagrams are included in the calculation of
the $\hat{Q}$-box, while the red bands are obtained if $U$-insertion
diagrams are calculated through third order in perturbation theory.

We observe a substantial reduction of the dependence on the choice of the
HO parameter as higher-order contributions of the $U$-insertion
diagrams are included, in particular the closure properties at $N=28$
are very sensitive to this issue.

As mentioned before, the $H_{\rm eff}$ derived for one-valence nucleon
systems contains only one-body contributions and provides the SP
energies for the SM calculation, while the two-body matrix elements
are obtained from $H_{\rm eff}$ derived from the two-valence nucleon
systems, once the theoretical SP energies are subtracted from its
diagonal matrix elements.

In order to perform our study, we have derived for each nucleus two
classes of $H_{\rm eff}$s; one has been obtained calculating
$\hat{Q}$-box diagrams with 2NF vertices only, dubbed as $H_{\rm
  eff}^{\rm 2NF}$.
The other, indicated as $H_{\rm eff}^{\rm 3NF}$, has been built up
including also $H^{\rm 3NF}_1$ first-order contributions in the
collection of $\hat{Q}$-box diagrams (see Fig. \ref{1b2b3bf}).
In the Supplemental Material \cite{supplemental2019} the TBMEs of
$H_{\rm eff}^{\rm 2NF}, H_{\rm eff}^{\rm 3NF}$ for systems with two
valence nucleons only can be found, while the proton and neutron SP
energies calculated with respect to $0f_{7/2}$ orbital -
$\epsilon_{\pi}$ and $\epsilon_{\nu}$ respectively - are reported in
Table \ref{tablespe}.

\begin{table}[ht]
\caption{Theoretical proton and neutron SP energies (in
  MeV) from $H_{\rm eff}^{\rm 2NF}$ and $H_{\rm eff}^{\rm 3NF}$.}
\begin{ruledtabular}
  \begin{tabular}{cccccccc}
 ~&~& $H_{\rm eff}^{\rm 2NF}$ &~&~&~& $H_{\rm eff}^{\rm 3NF}$&~\\
  ~ & $\epsilon_{\pi}$ & ~ & $\epsilon_{\nu}$ & ~ & $\epsilon_{\pi}$ &
                                                                       ~
                                                             &
                                                               $\epsilon_{\nu}$\\
    \colrule
    ~ & ~ & ~ & ~ & ~ & ~ & ~ & ~\\
$0f_{7/2}$   & 0.0 & ~ & 0.0 & ~ & 0.0 & ~ & 0.0 \\ 
 $0f_{5/2}$   & 4.2 & ~ & 5.1 & ~ & 5.5 & ~ & 7.4 \\ 
 $1p_{3/2}$   & 0.0 & ~ & 0.5 & ~ & 1.6 & ~ & 2.8 \\ 
 $1p_{1/2}$   & 1.0 & ~ & 2.0 & ~ & 2.9 & ~ & 4.9 \\ 
\end{tabular}
\end{ruledtabular}
\label{tablespe}
\end{table}

In  order to accomplish our goal to investigate the shell evolution of
spectroscopic properties of $fp$ nuclei, we have performed a multipole
decomposition of $H_{\rm eff}^{\rm 2NF}$ and $H_{\rm eff}^{\rm 3NF}$
for any isotope under investigation \cite{Umeya08,Umeya16}, focussing
our interest on their monopole components.
It is worth recalling that the angular-momentum-averaged monopole
component of the shell-model Hamiltonian is defined as follows:

\begin{eqnarray}
\langle i,j | H_{\rm eff}^{mon} | i,j \rangle & = & \epsilon_i + \epsilon_j+
\frac{\sum_J  (2J +1) \langle i,j |V_{\rm eff} | i,j \rangle_J}{
                                                    \sum_J (2J +1)}=
                                                    \nonumber \\
~ & = & \epsilon_i + \epsilon_j+ V^{mon}_{ij}~~,
\label{eqmon}
\end{eqnarray}

\noindent
where $V_{\rm eff}$ is the two-body component of $H_{\rm eff}$, $i$
and $j$ indicate the quantum numbers of the SP states,
and the $\epsilon_i$ are the SP energies.
Consequently, we have also studied the evolution of the proton and
neutron ESPEs as a function of the valence nucleons, that are defined as:

\begin{equation}
  {\rm ESPE}(j) = \epsilon_j + \sum_{j'} V^{mon}_{j j'} n_{j'}~~,
\label{eqespe}
\end{equation}

\noindent
where the sum runs over the model-space levels $j'$, $n_{j}$ being the number
of particles in the level $j$.

\section{Results}
\label{results}
\subsection{Monopole components of the effective SM Hamiltonians}
\begin{figure}[h]
\begin{center}
\includegraphics[scale=0.30,angle=0]{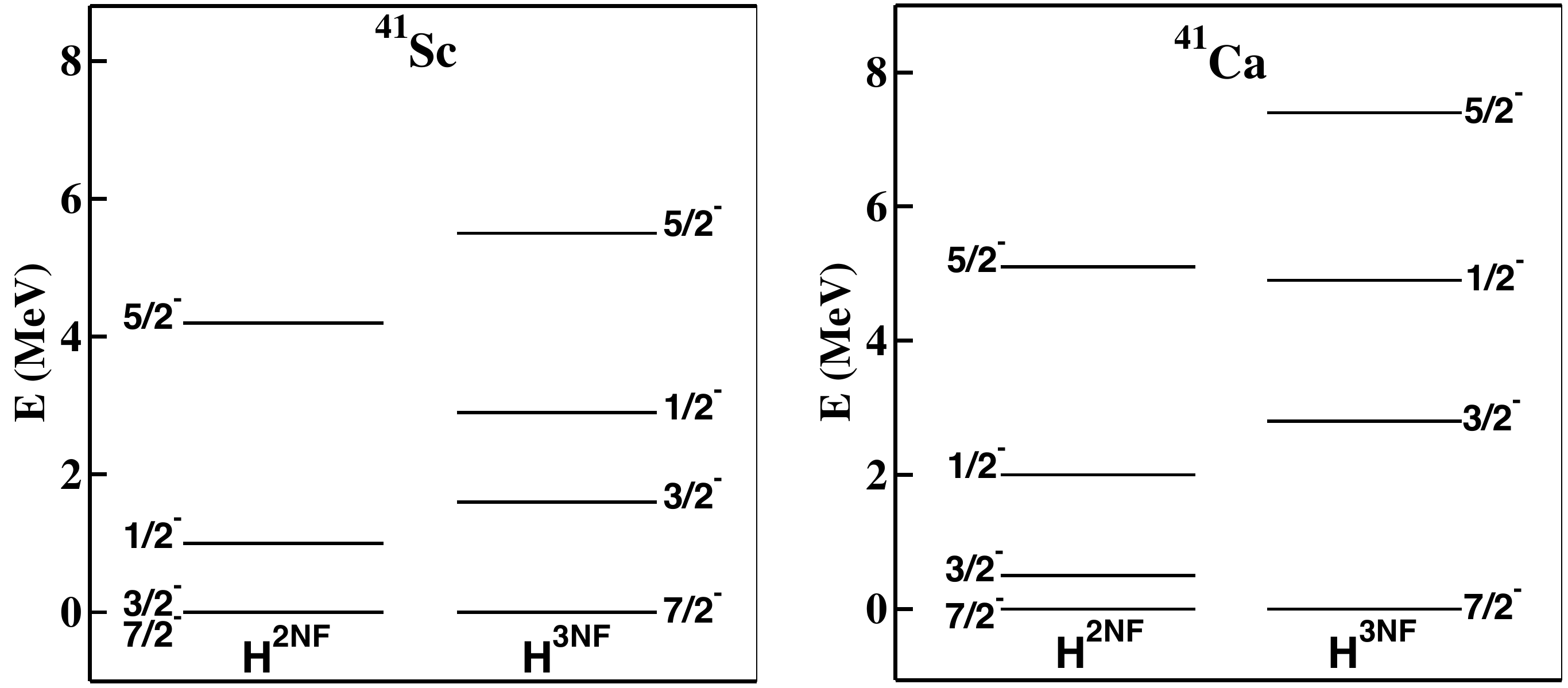}
\caption{Calculated SP spectra of $^{41}$Sc and $^{41}$Ca, as
  obtained from $H_{\rm eff}^{\rm 2NF}$ and $H_{\rm eff}^{\rm
  3NF}$. They represent the proton and neutron SP energies,
respectively, employed in our calculations.}
\label{41Sc41Ca}
\end{center}
\end{figure}

Before we start our discussion about the characteristics of the
monopole component of $H_{\rm eff}^{\rm 2NF}$ and $H_{\rm eff}^{\rm
  3NF}$, it is worth coming back to the calculated SP energies of
both effective Hamiltonians, which can be found in Fig. \ref{41Sc41Ca}
as single-particle spectra of $^{41}$Sc and $^{41}$Ca.
We do not show in this figure any experimental counterpart, because
the experimental information about the spectroscopic factors of both
nuclei are rather scanty, and consequently we have no clear
indications on the SP nature of the observed low-energy levels
\cite{ensdf}.

From the inspection of Fig. \ref{41Sc41Ca}, we observe that $H_{\rm
  eff}^{\rm 2NF}$ does not provide enough spin-orbit splitting between
the $0f_{7/2,5/2}$ orbitals in both $^{41}$Sc and $^{41}$Ca.
Moreover, the $0f_{7/2}$ and $1p_{3/2}$ orbitals are not
well-separated and, consequently, it can be inferred that calculations
with $H_{\rm eff}^{\rm 2NF}$ might not be able to describe the shell
closure that is observed at $Z,N=28$.
On the other hand, the contribution coming from the 3NF is able to
heal this defect of the SM Hamiltonian, and in the SP spectrum of
$H_{\rm eff}^{\rm 3NF}$ the $0f_{7/2}$ orbital is lowered enough with
respect to the $1p_{3/2},1p_{1/2},0f_{5/2}$ orbitals to lay the
foundation of a better shell closure at $N,Z=$28.

\begin{figure}[h]
\begin{center}
\includegraphics[scale=0.34,angle=0]{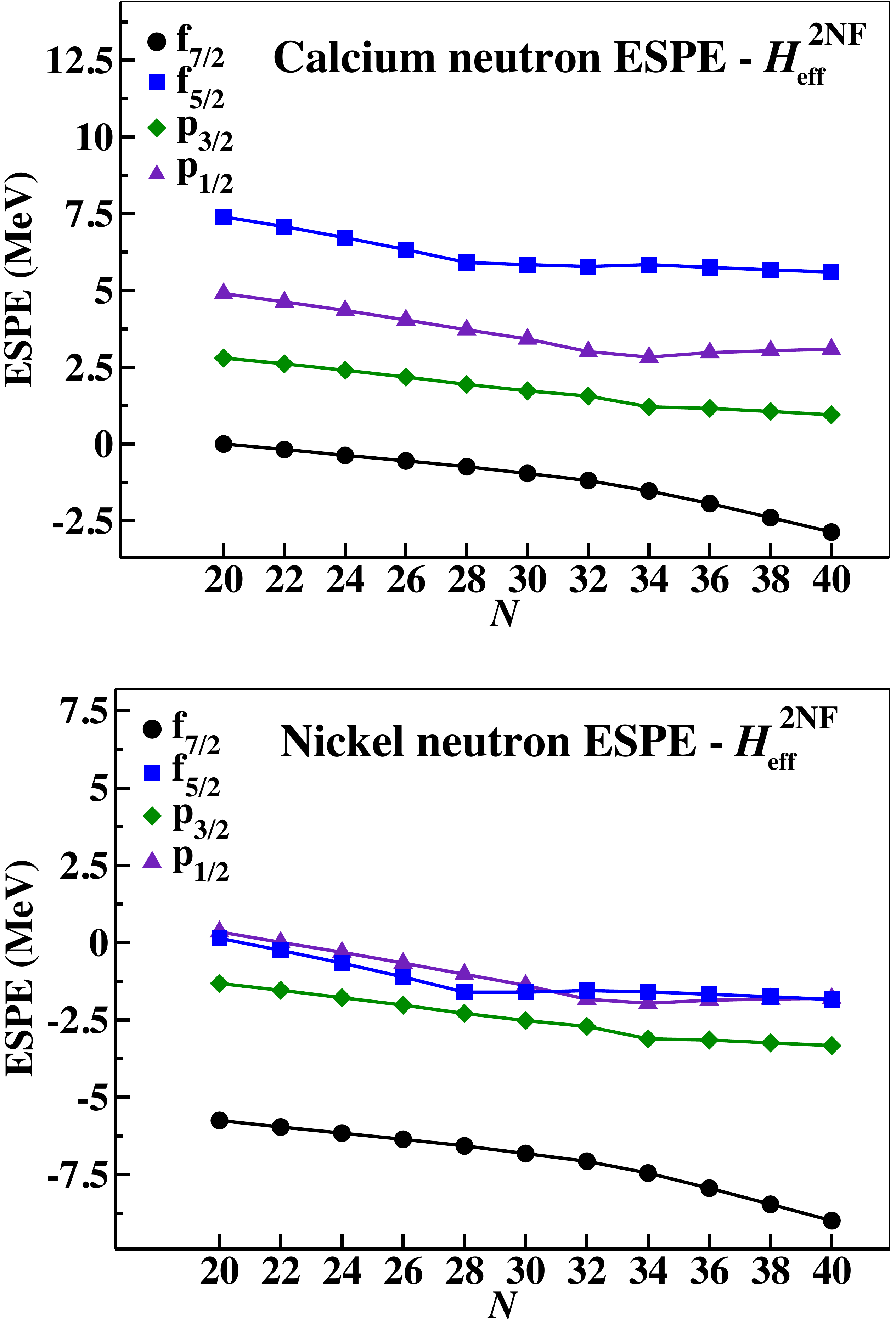}
\caption{Neutron ESPEs from $H^{\rm 2NF}_{\rm eff}$ TBMEs for calcium
  and nickel isotopes as a function of the neutron number (see
  the text for details).}
\label{espe_2nf_nn}
\end{center}
\end{figure}

Actually, a shell closure cannot be guaranteed only by the SP energy
spacings, since the TBMEs of $H_{\rm eff}$ play a crucial role in their
evolution as a function of the valence-nucleon number.
As a matter of fact, in Ref. \cite{Coraggio18a} the SP energies of
$p$-shell nuclei, calculated with and without 3NF contributions, start
both from a sufficient spin-orbit splitting to provide, in principle,
the $Z,N=6$ subshell closure.
However, we have found that the monopole component of $H^{\rm
  2NF}_{\rm eff}$ compresses the separation between the $0p_{3/2}$ and
$0p_{1/2}$ orbitals when increasing the valence-nucleon number, at
variance with the $H^{\rm 3NF}_{\rm eff}$ monopole term that preserves
a constant energy spacing.

On the above ground, a study of the evolution of the ESPEs of $H^{\rm
  2NF}_{\rm eff}$ and $H^{\rm 3NF}_{\rm eff}$ in terms of the
valence-nucleon number is highly desirable to understand how to obtain
a sound description of their shell closure properties.
This evolution of the ESPEs depends only on the TBMEs, and in the
following we decide to report the neutron ESPEs of calcium isotopes
and both neutron and proton ESPEs of nickel isotopes, as a function
of the number of valence neutrons, calculated employing the TBMEs of
$H^{\rm 2NF}_{\rm eff}$ and $H^{\rm 3NF}_{\rm eff}$, but starting from
the same set of SP energies, namely those of $H^{\rm 3NF}_{\rm eff}$.
This is done to evidence the relevant features of $H^{\rm 2NF}_{\rm
  eff},H^{\rm 3NF}_{\rm eff}$ monopole components, and to infer their
different shell-evolution properties around doubly-closed $^{48}$Ca
and $^{56}$Ni.

Figure \ref{espe_2nf_nn} shows the neutron ESPEs of
  calcium and nickel isotopes obtained with $H^{\rm 2NF}_{\rm eff}$
  TBMEs, starting from $H^{\rm 3NF}_{\rm eff}$ SP energies, and
  evolved as a function of the valence neutrons up to $N=40$.
Black dots, blue squares, green diamonds, and indigo triangles
indicate the $0f_{7/2}$, $0f_{5/2}$, $1p_{3/2}$, and $1p_{1/2}$ ESPE,
respectively. 

As can be seen, the spacings between the $fp$ orbitals
  remain almost constant with respect to the evolution of the
  valence-neutron number, with the $0f_{7/2}$ ESPE well separated from
  the other ones.
  For the calcium isotopes also $1p_{1/2}$, $0f_{5/2}$ orbitals are
  separated from the $1p_{3/2}$ one and between themselves too, while
  neutron ESPEs of nickel isotopes reveal that these three orbitals
  are grouped and very close to each other.
  
This feature seems to point to a reasonable shell closure in
doubly-closed $^{48}$Ca when employing the $H^{\rm 3NF}_{\rm eff}$
neutron SP spacings reported in Table \ref{tablespe} and TBME obtained
from $H^{\rm 2NF}_{\rm eff}$, and also to a pronounced subshell
closure at $N=32$ and $N=34$ for calcium isotopes.
This is consistent with the results we obtained in a previous work
\cite{Coraggio09c}, whose focus was the study of the spectroscopic
properties of neutron-rich calcium isotopes. 
In that paper, the TBME were extracted from a $H_{\rm eff}^{V_{\rm
    low-k}}$ derived from the CD-bonn potential \cite{Machleidt01b}
renormalized by way of the $V_{\rm low-k}$ procedure, while the SP
energies were fitted on experimental SP states in $^{47,49}$Ca.
As a matter of fact, the role of three-body forces is mainly absorbed
by the procedure of fixing SP energies to reproduce SP observables;
actually, in a recent paper \cite{Coraggio19a} we have shown that the
theoretical SP energies obtained from $H_{\rm eff}^{V_{\rm low-k}}$ do
not reproduce the observed shell-closure of the neutron $0f_{7/2}$
orbital in $^{48}$Ca, the agreement between the experimental and
calculated spectra of this nucleus being only qualitative.

  As regards the nickel isotopes, the close values of $1p_{3/2}$,
  $1p_{1/2}$, $0f_{5/2}$ ESPEs may influence the shell closure in
  $^{56}$Ni and provide the disappearance of $N=32$ and $N=34$
  subshell closures.

\begin{figure}[h]
\begin{center}
\includegraphics[scale=0.34,angle=0]{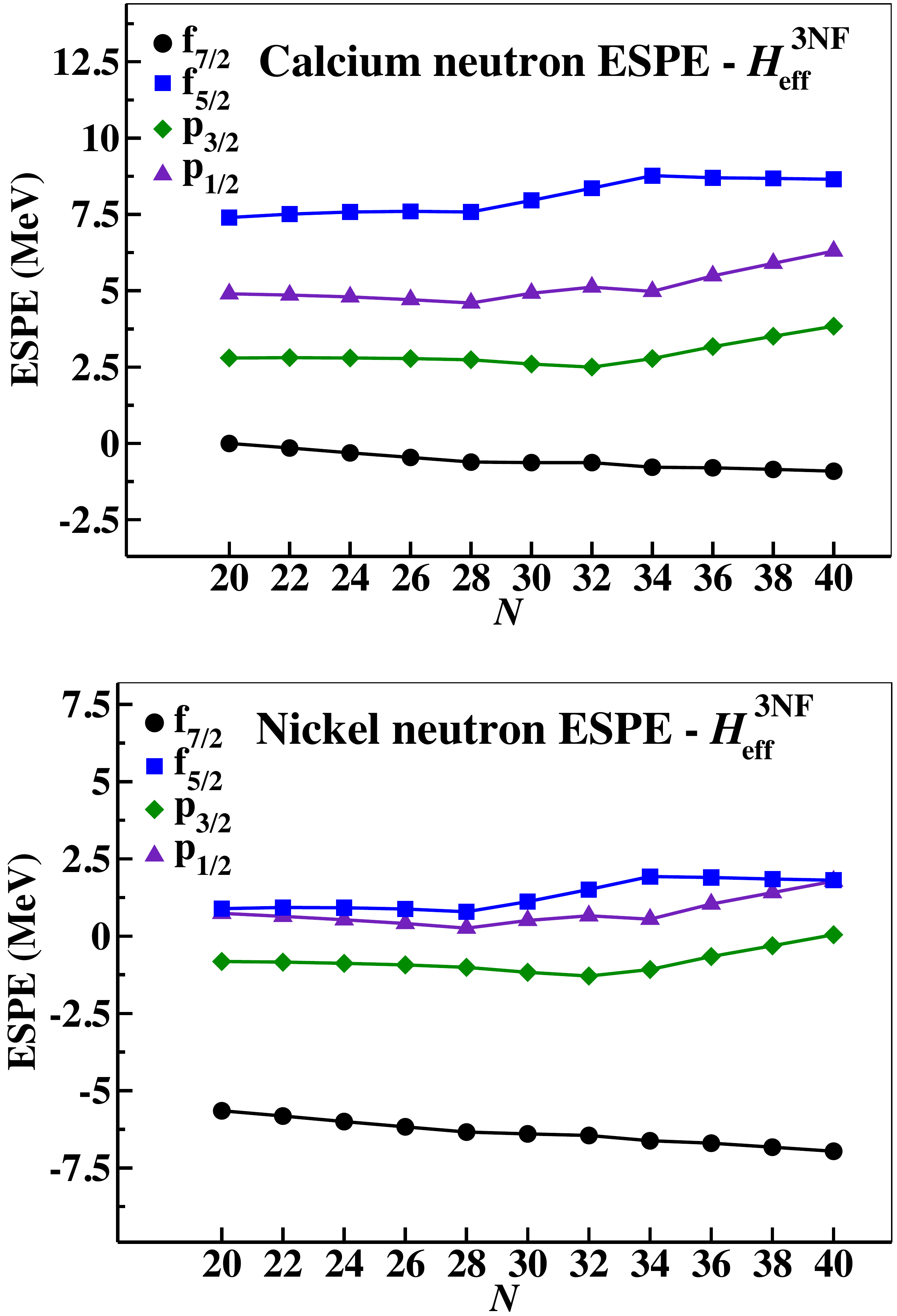}
\caption{Same as in Fig. \ref{espe_2nf_nn}, but for $H^{\rm 3NF}_{\rm
    eff}$.}
\label{espe_3nf_nn}
\end{center}
\end{figure}

The neutron ESPEs obtained from $H^{\rm 3NF}_{\rm eff}$ TBME  are
presented in Fig. \ref{espe_3nf_nn} for both calcium and nickel
isotopes.

The inclusion of 3NF effects does not affect the general behavior of
the neutron ESPEs for both isotopic chains, but some specific details
may reveal relevant features that will show up in the results of the
full SM calculations in the next section.

As regards the calcium isotopes, at $N=28$ the neutron monopole
component of $H^{\rm 3NF}_{\rm eff}$ enlarges the $0f_{7/2} -
1p_{3/2}$ gap by 0.7 MeV, inducing a stronger shell closure.
Also the $1p_{1/2}- 1p_{3/2}$ and $0f_{5/2}- 1p_{1/2}$ splittings at
$N=32$ and $N=34$, respectively, grow and strengthen the corresponding
subshell closures, as we will show in the next section.

The 3NF contribution to the neutron ESPEs provides also a stronger
closure in $^{56}$Ni since the gap between $1p_{3/2}$ and $0f_{7/2}$
orbitals at $N=28$ is 1 MeV larger than the one reported in
Fig. \ref{espe_2nf_nn}, that is calculated with TBME obtained from
$H^{\rm 2NF}_{\rm eff}$.

\begin{figure}[h]
\begin{center}
\includegraphics[scale=0.33,angle=0]{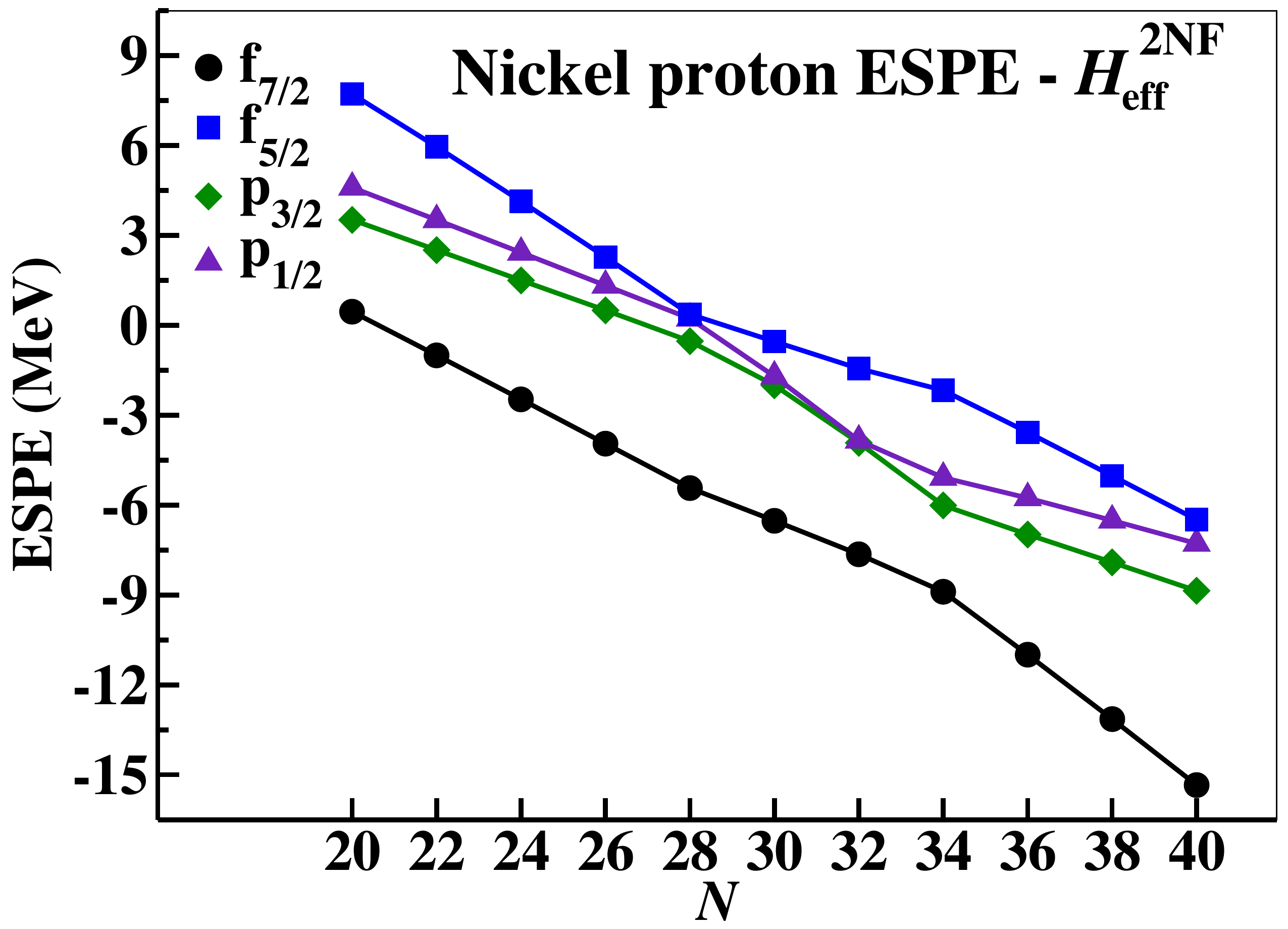}
\caption{Proton ESPEs from $H^{\rm 2NF}_{\rm eff}$ TBMEs for nickel
  isotopes as a function of the neutron number.}
\label{espe_2nf_pn}
\end{center}
\end{figure}

The above considerations about the $^{56}$Ni shell closure are
strengthened if we consider also the evolution of proton ESPEs of
nickel isotopes as a function of the valence-neutron number.

\begin{figure}[H]
\begin{center}
\includegraphics[scale=0.33,angle=0]{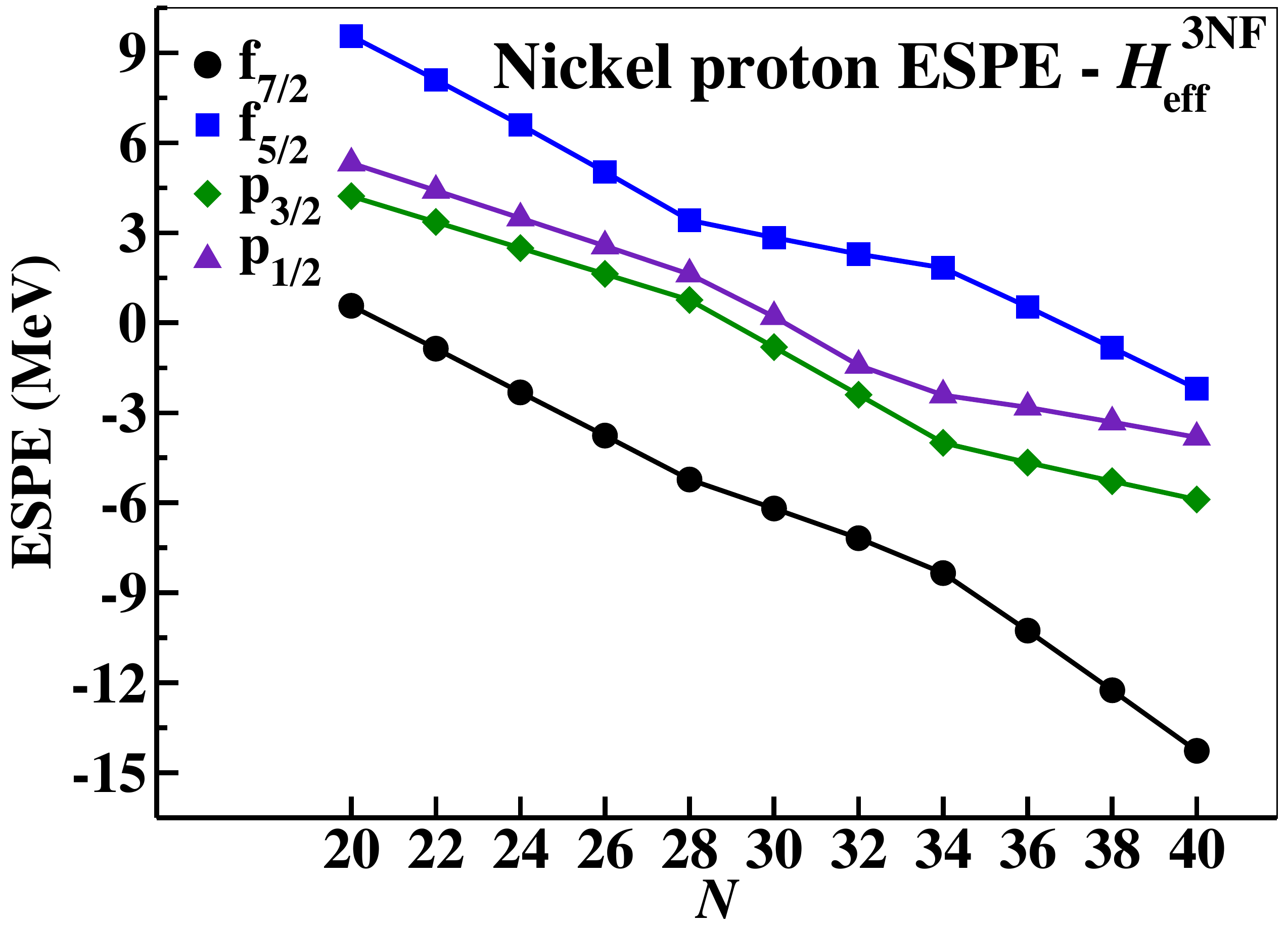}
\caption{Same as in Fig. \ref{espe_2nf_pn}, but for $H^{\rm 3NF}_{\rm
    eff}$ proton ESPEs.}
\label{espe_3nf_pn}
\end{center}
\end{figure}

As can be seen in Figs. \ref{espe_2nf_pn}, \ref{espe_3nf_pn}, the
separation in energy between the $0f_{5/2}$ and $0f_{7/2}$ ESPEs is about
5.8 MeV and 8.6 MeV at $N=28$, calculated with $H^{\rm 2NF}_{\rm eff}$
and $H^{\rm 3NF}_{\rm eff}$, respectively.
Moreover, the gap between the proton ESPEs of $0f_{5/2}$ and
$1p_{3/2}$ orbitals reduces to 0.8 MeV at $N=28$,  if only 2NF is
considered to derive the shell-model effective Hamiltonian, while the
3NF contributions limit this reduction to 1.6 MeV.

These features should induce a collective effect at $N=28$, and a less
pronounced shell-closure for $^{56}$Ni than $^{48}$Ca.
This collectivity affects the results obtained with $H^{\rm
  2NF}_{\rm eff}$ more than those with $H^{\rm 3NF}_{\rm eff}$, as we
will see in the next section.

\subsection{Shell-model calculations}
\label{smcalculations}

There are some spectroscopic features which reveal the shell closure
properties, and among them two of the most important ones are the
behavior of the excitation energy of $J^{\pi}_1=2^+$ states and the
evolution of the ground-state (g.s.) energy in even mass
isotopic/isotonic chains, with respect to the number of valence
neutrons/protons.

\begin{figure}[h]
\begin{center}
\includegraphics[scale=0.31,angle=0]{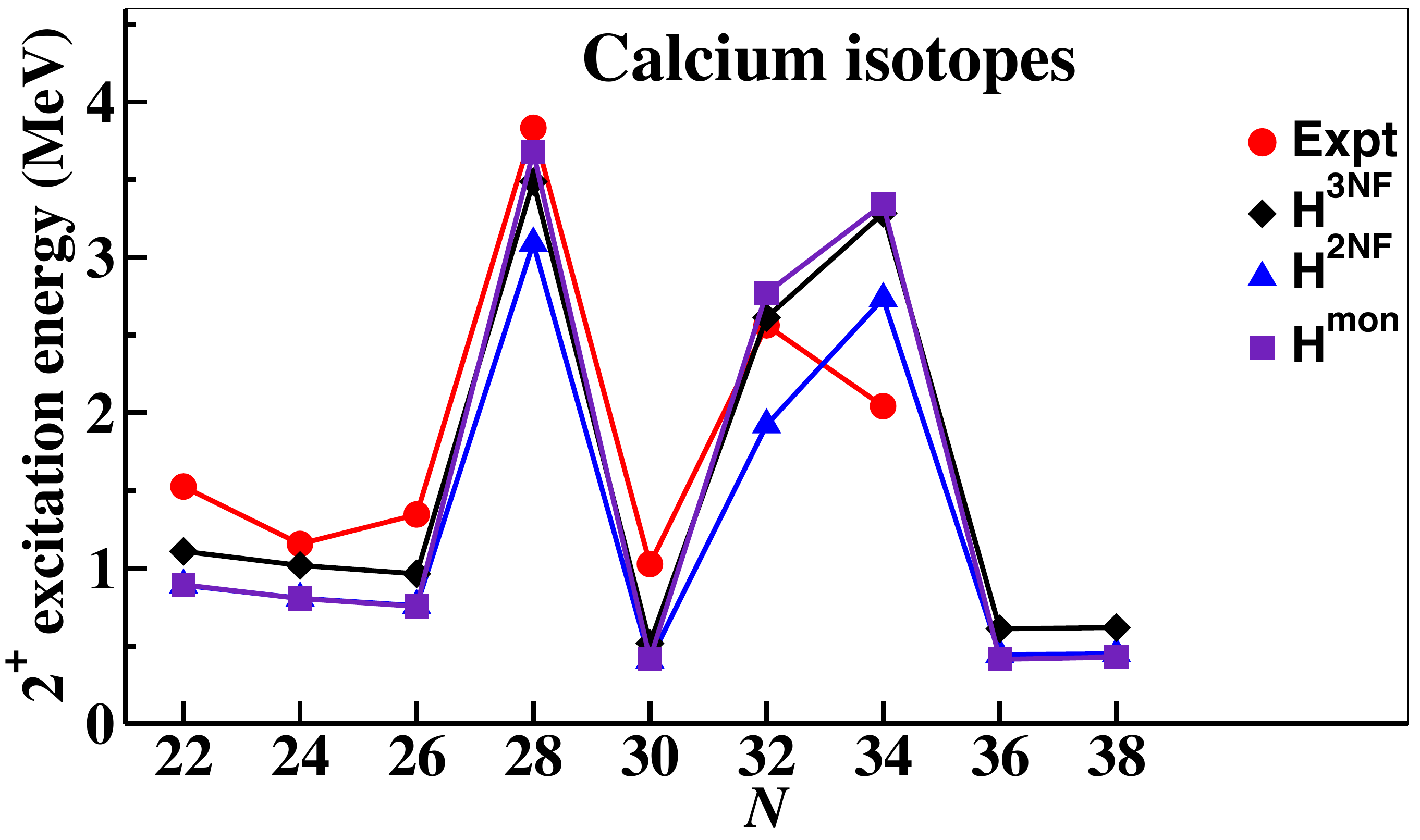}
\caption{Experimental and calculated excitation energies of the yrast
  $J^{\pi}=2^+$  states for calcium isotopes from $N = 22$ to
  38. See text for details.}
\label{J2pCa}
\end{center}
\end{figure}

These properties will be investigated by diagonalizing the two classes
of Hamiltonians $H^{\rm 2NF}_{\rm eff}$ and $H^{\rm 3NF}_{\rm eff}$,
and employing for both of them the set of SP energies provided by
$H^{\rm 3NF}_{\rm eff}$.
We refer to {\it class of effective Hamiltonians} since, as reported
in Sec. \ref{outline}, they change according to the number of valence
protons and neutrons because of the density dependence introduced by
accounting for three-body correlation diagrams.

In addition to these two classes of effective SM Hamiltonians, we have
built another one, that we dub $H^{\rm mon}_{\rm eff}$, by summing the
monopole component of $H^{\rm 3NF}_{\rm eff}$ and the multipole ones
belonging to $H^{\rm 2NF}_{\rm eff}$.
The scope of this operation is to evidence the interplay of the
monopole and multipole components through the diagonalization of the
effective SM Hamiltonian, and will be better clarified in the
discussion of the result of our calculations.

The experimental and theoretical results obtained with $H^{\rm
  2NF}_{\rm eff}$, $H^{\rm 3NF}_{\rm eff}$, and $H^{\rm mon}_{\rm
  eff}$ will be indicated in the figures with red dots, blue triangles,
black diamonds, and indigo squares, respectively.

We start our study with calcium isotopes, and in Fig. \ref{J2pCa}
they are shown the $J^{\pi}=2^+_1$ excitation energies from $N=22$ up
to $N=38$.

We observe that the results obtained with all three Hamiltonians are
very similar.
The shell closure at $N=28$ is very-well reproduced by $H^{\rm
  3NF}_{\rm eff}$ and $H^{\rm mon}_{\rm eff}$, while the $J^{\pi}=2^+_1$ excitation
energy obtained with $H^{\rm 2NF}_{\rm eff}$ is about 0.7 MeV lower
than the experimental one \cite{ensdf}.
The different results for the $^{48}$Ca shell-closure trace back to
the different energy gap between the $1p_{3/2}$ and $0f_{7/2}$ neutron
ESPE when we employ the monopole term of $H^{\rm 2NF}_{\rm eff}$ and
$H^{\rm 3NF}_{\rm eff}$, as can be seen in Figs. \ref{espe_2nf_nn} and
\ref{espe_3nf_nn}.

There are present also two subshell closures at $N=32,34$, the second
one being too strong when compared with experiment.
As a matter of fact, a preliminary study of calcium isotopes,
performed with a larger model space that includes the $0g_{9/2}$
orbital too, shows that this enlargement of the model space is
mandatory to reproduce the observed behavior at $N=32,34$
\cite{Coraggio19c}.

\begin{figure}[H]
\begin{center}
\includegraphics[scale=0.31,angle=0]{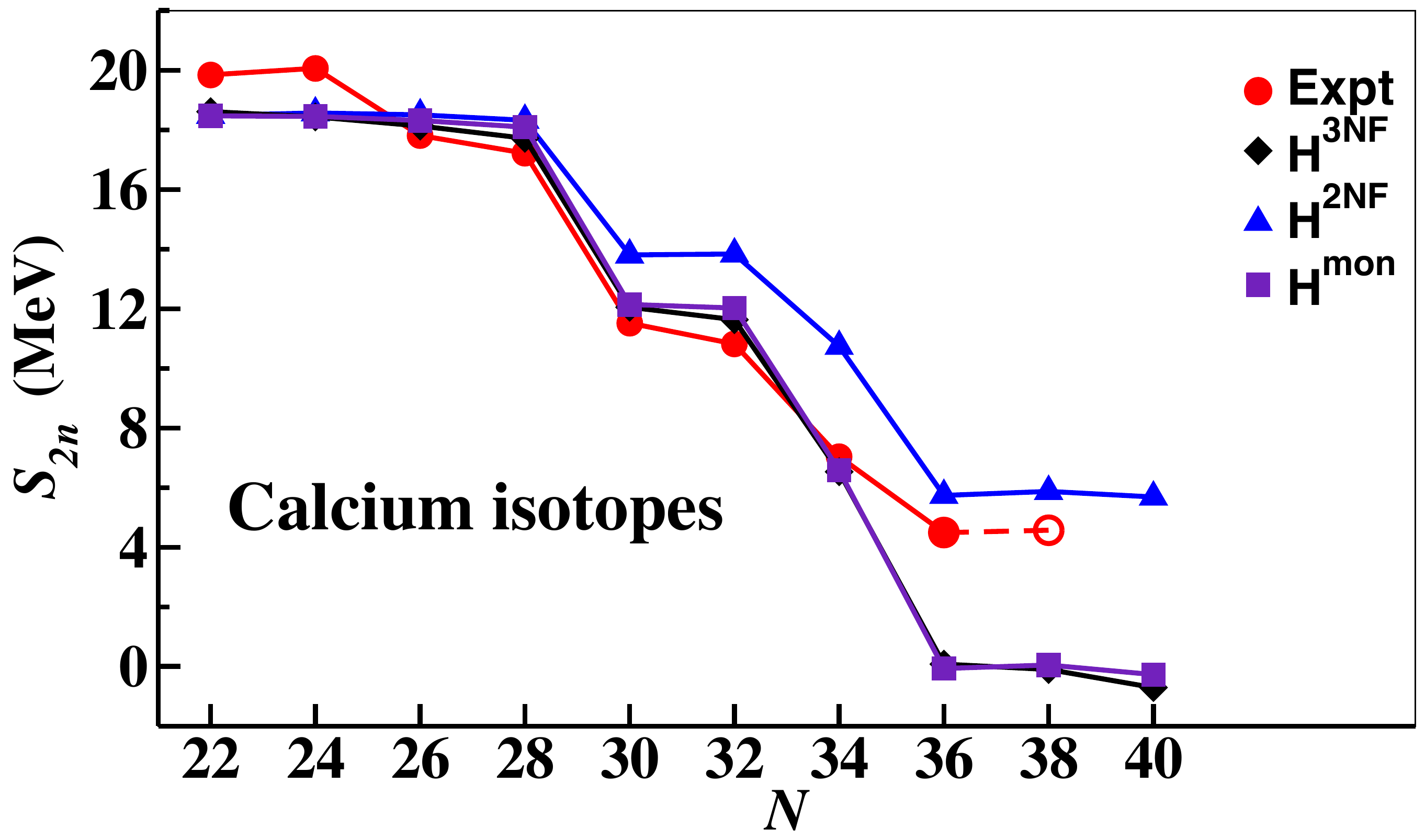}
\caption{Experimental and calculated two-neutron separation energies
  for calcium isotopes from $N = 22$ to 40. Data are taken from
 \cite{Audi03,Wienholtz13,Michimasa18}, open circles correspond to
  estimated values reported in Ref. \cite{Audi03}. See text for details.}
\label{S2nCa}
\end{center}
\end{figure}

Different closure properties, related to whether 3NF are
included or not in the derivation of the effective SM Hamiltonian, are
present also in the calculation of the $S_{2n}$ that are shown in
Fig. \ref{S2nCa} for the calcium isotopes up to $N=40$.
As already mentioned in the previous section, we have
shifted the SP energies in Table \ref{tablespe} in order to reproduce
the experimental g.s. energy of $^{41}$Ca and $^{41}$Sc with respect
to $^{40}$Ca.

As can be seen, both experimental
\cite{Audi03,Wienholtz13,Michimasa18} and theoretical $S_{2n}$ show a
rather flat behavior up to $N=28$, then a sudden drop occurs at $N=30$
that is a signature of the shell closure due to the $0f_{7/2}$ filling. 
Another decrease appears at $N=34$ because at that point the valence
neutrons start to occupy the $1p_{1/2}$ and $0f_{5/2}$ orbitals.

It should be recalled that recently the comparison
  between experimental and calculated masses at $N=32,34$ of
  neutron-rich calcium isotopes has been spotted as a way to pin down
  the role of 3NF in nuclear structure calculations
  \cite{Wienholtz13,Holt14}.

The results obtained with $H^{\rm 3NF}_{\rm eff}$ and $H^{\rm
  mon}_{\rm eff}$ follow closely the behavior of the experimental
$S_{2n}$ up to $N=34$, while those obtained with $H^{\rm 2NF}_{\rm
  eff}$ provide a less satisfactory energy drop between $N=28$ and 30.

At $N=36$, the repulsive 3NF effects contribute to a sudden drop of
the two-neutron separation energies, in contrast with the experimental
values.
As for the case of the calculated yrast $J^{\pi}=2^+$ excitation
energies, we need to point out that a larger model space, including at
least the $0g_{9/2}$ orbital, improves the depiction of the
spectroscopic properties of heavy-calcium isotopes
\cite{Coraggio19c}.
Within such an enlarged model space, we have found that $H^{\rm
  3NF}_{\rm eff}$ and $H^{\rm mon}_{\rm eff}$ provide a limit of the
neutron dripline that is consistent with the recent observation of a
bound $^{60}$Ca \cite{Tarasov18}, while from the inspection of
Fig. \ref{S2nCa} we observe that present results predict the calcium
dripline located at $N=38$.

Now we move from systems with identical valence particle to those with
both valence protons and neutrons, in order to investigate the changes
in the shell evolution and closure properties originating from the
collectivity ignited by the $T=0$ channel of the residual interaction.

\begin{figure}[H]
\begin{center}
\includegraphics[scale=0.31,angle=0]{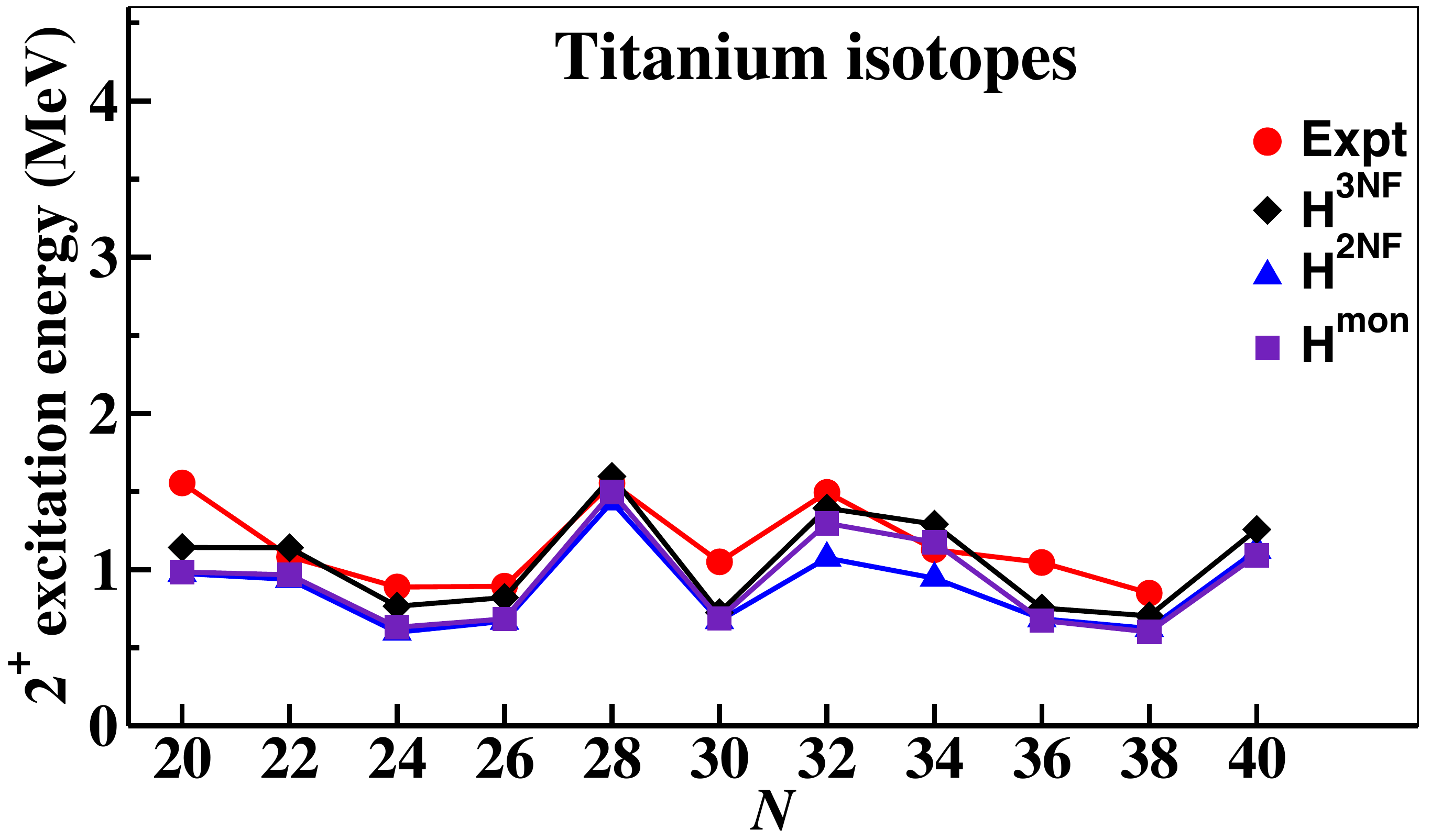}
\caption{ Same as in Fig. \ref{J2pCa}, but for titanium isotopes from
  $N = 20$ to 40. See text for details.}
\label{J2pTi}
\end{center}
\end{figure}

In Fig. \ref{J2pTi} the calculated $J^{\pi}=2^+_1$ excitation energies
of titanium isotopes are reported and compared with data
\cite{ensdf}.
We observe that the experimental behavior is, overall, well reproduced
by all three SM Hamiltonians up to $N=34$, the largest
discrepancies occurring for $^{42}$Ti and $^{52}$Ti with all effective
Hamiltonians, and for $^{54}$Ti with $H^{\rm 2NF}_{\rm eff}$.

As regards the results for heavier isotopes, the underestimation of
the experimental results points to the need to employ a larger model
space, as already mentioned.

\begin{figure}[H]
\begin{center}
\includegraphics[scale=0.31,angle=0]{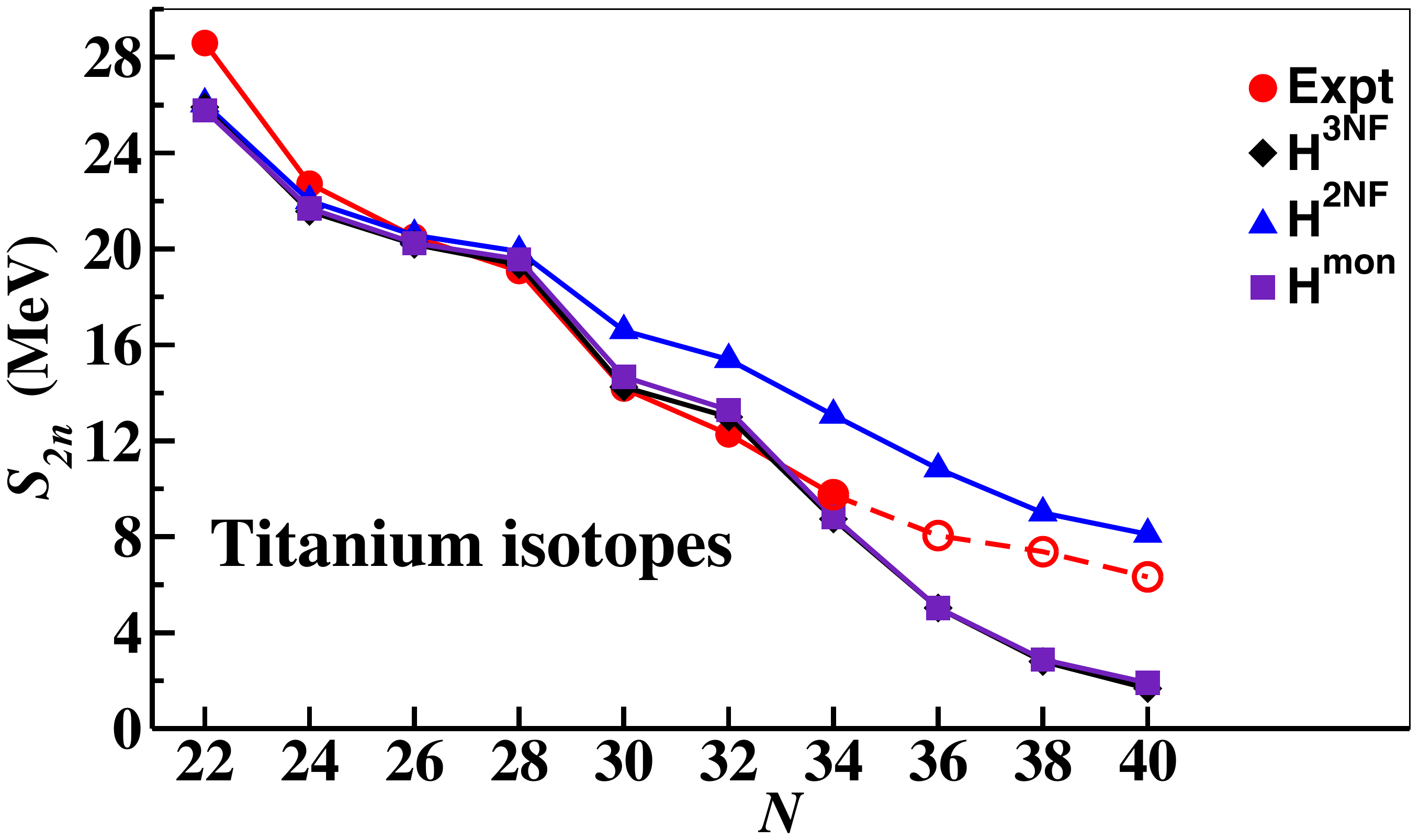}
\caption{Same as in Fig. \ref{S2nCa}, but for titanium isotopes from
  $N = 22$ to 40. See text for details.}
\label{S2nTi}
\end{center}
\end{figure}

From the inspection of Fig. \ref{S2nTi}, we observe that also the
$S_{2n}$ experimental behavior \cite{Audi03} is well reproduced by
$H^{\rm 3NF}_{\rm eff}$ and $H^{\rm mon}_{\rm eff}$, while the
calculations with $H^{\rm 2NF}_{\rm eff}$ underestimate the drop of
two-neutron separation energy between $N=28$ and 30.
The latter feature evidences that also when the $T=0$ channel is
involved, the contribution of 3NF helps to obtain a better comparison
with experiment up to $N=34$. 

\begin{figure}[H]
\begin{center}
\includegraphics[scale=0.31,angle=0]{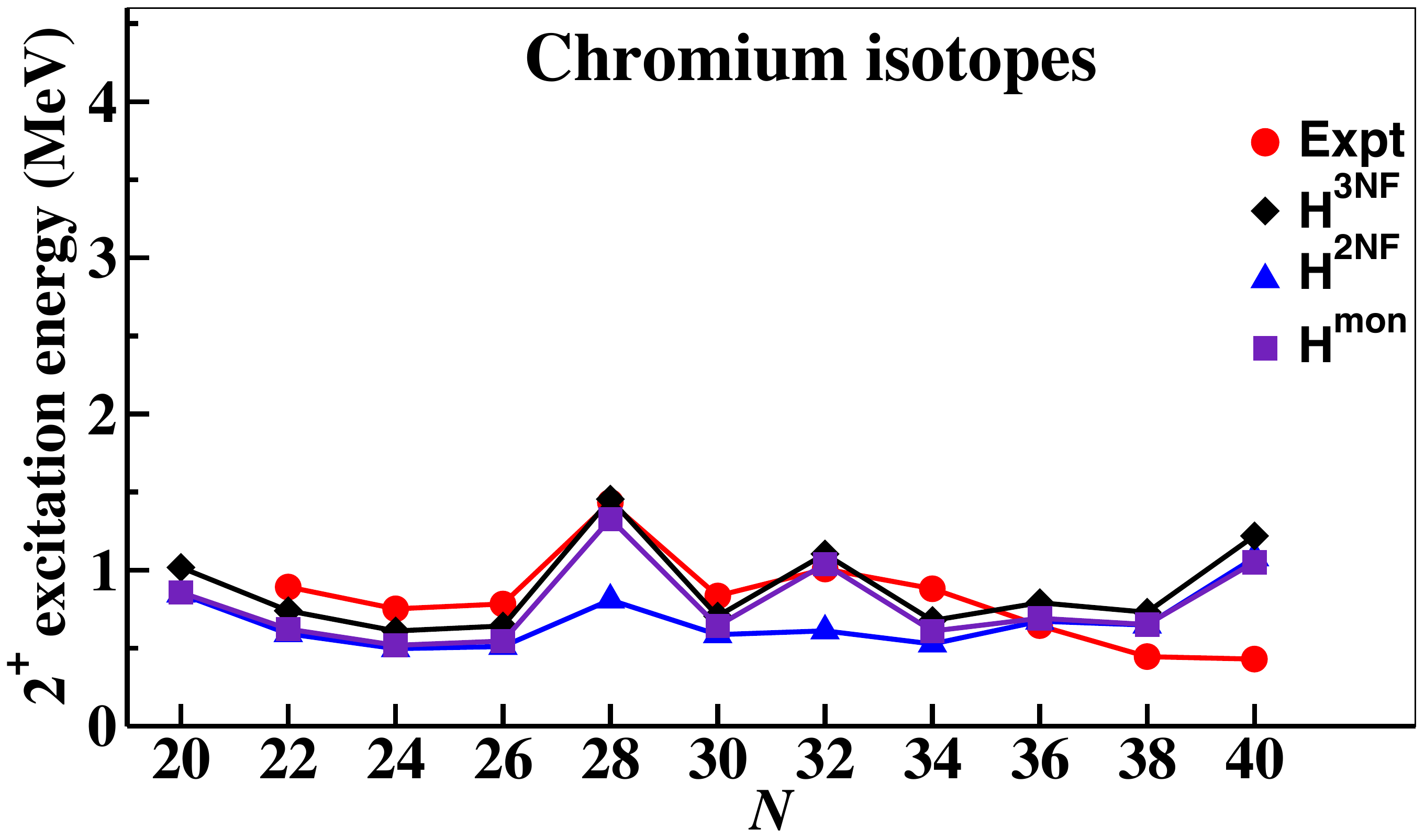}
\caption{ Same as in Fig. \ref{J2pTi}, but for chromium isotopes. See
  text for details.}
\label{J2pCr}
\end{center}
\end{figure}

\begin{figure}[H]
\begin{center}
\includegraphics[scale=0.31,angle=0]{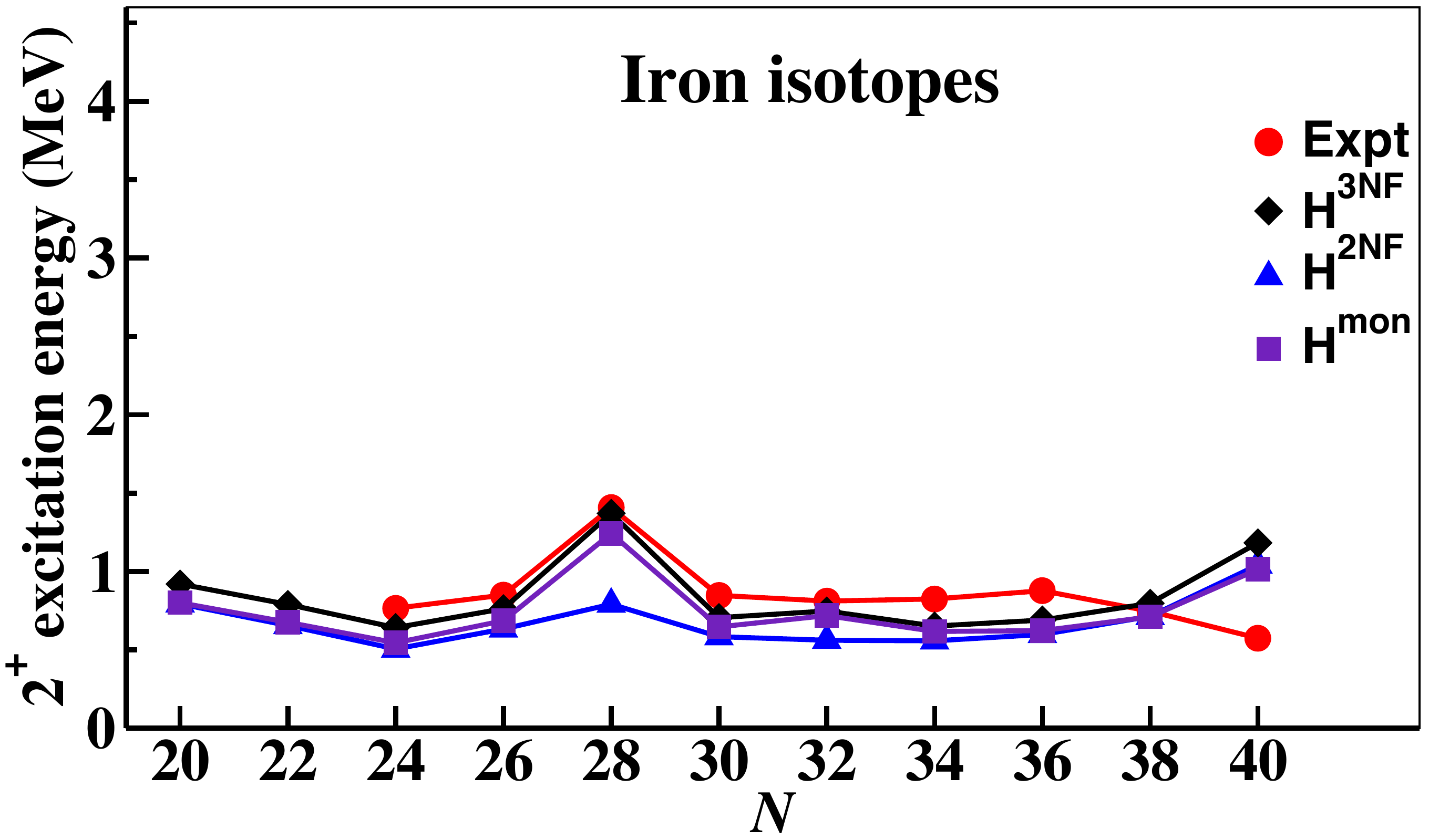}
\caption{ Same as in Fig. \ref{J2pTi}, but for iron isotopes. See text
  for details.}
\label{J2pFe}
\end{center}
\end{figure}

The collective behavior increases with the number of interacting
protons and neutrons, as can be observed for the chromium and iron
isotopes.
In Figs. \ref{J2pCr},\ref{J2pFe} we report the experimental
\cite{ensdf} and calculated excitation energies of the yrast
$J^{\pi}=2^+$ states up to $N=40$ for both isotopic chains.
We observe in both cases that the calculations with $H^{\rm 2NF}_{\rm
  eff}$ provide too much collectivity at $N=28$, while effective SM
Hamiltonians, whose monopole component includes 3NF contributions, are
able to reproduce the experimental behavior up to $N=34-36$ rather
well.

\begin{figure}[H]
\begin{center}
\includegraphics[scale=0.31,angle=0]{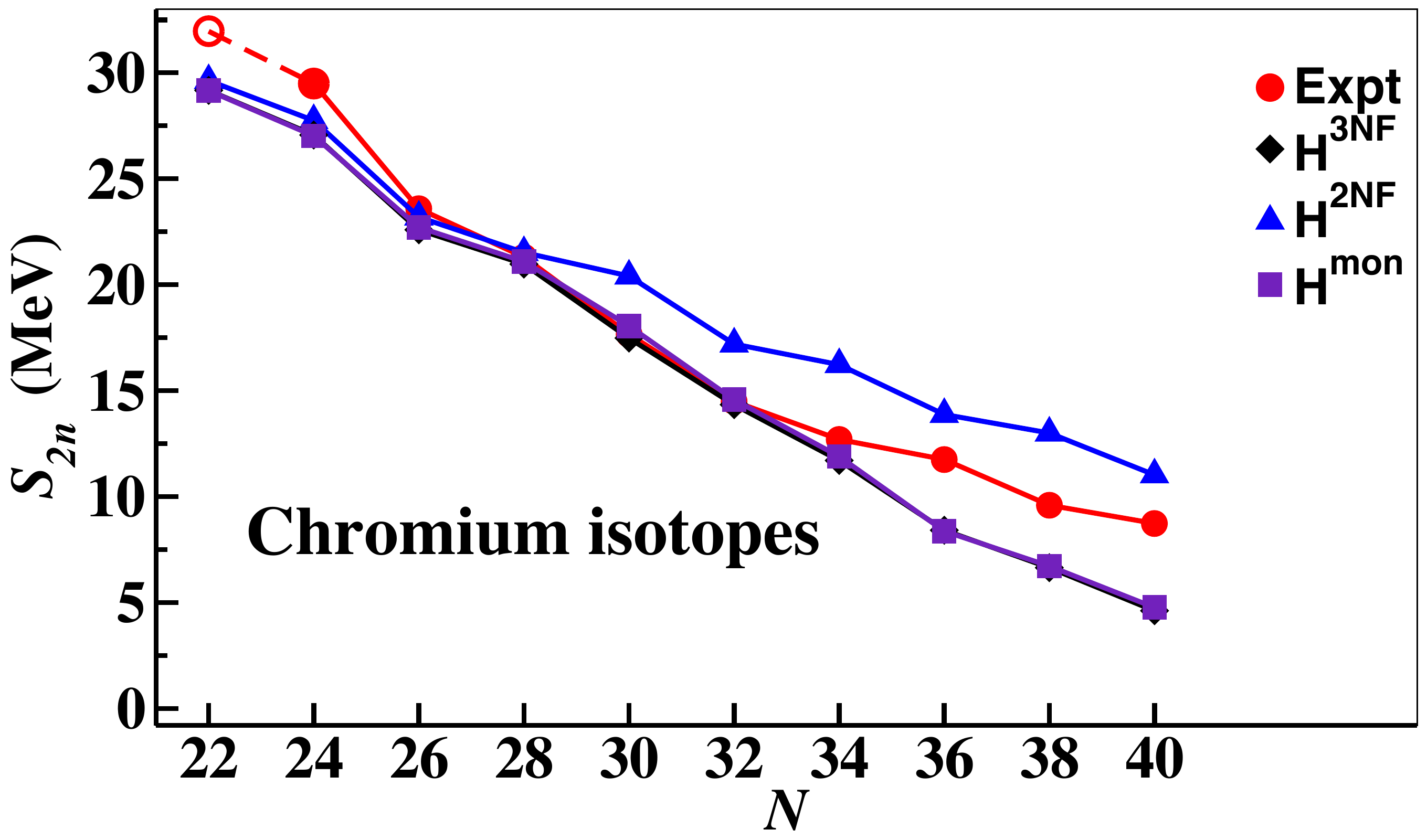}
\caption{Same as in Fig. \ref{S2nTi}, but for chromium isotopes. See
  text for details.}
\label{S2nCr}
\end{center}
\end{figure}

\begin{figure}[H]
\begin{center}
\includegraphics[scale=0.31,angle=0]{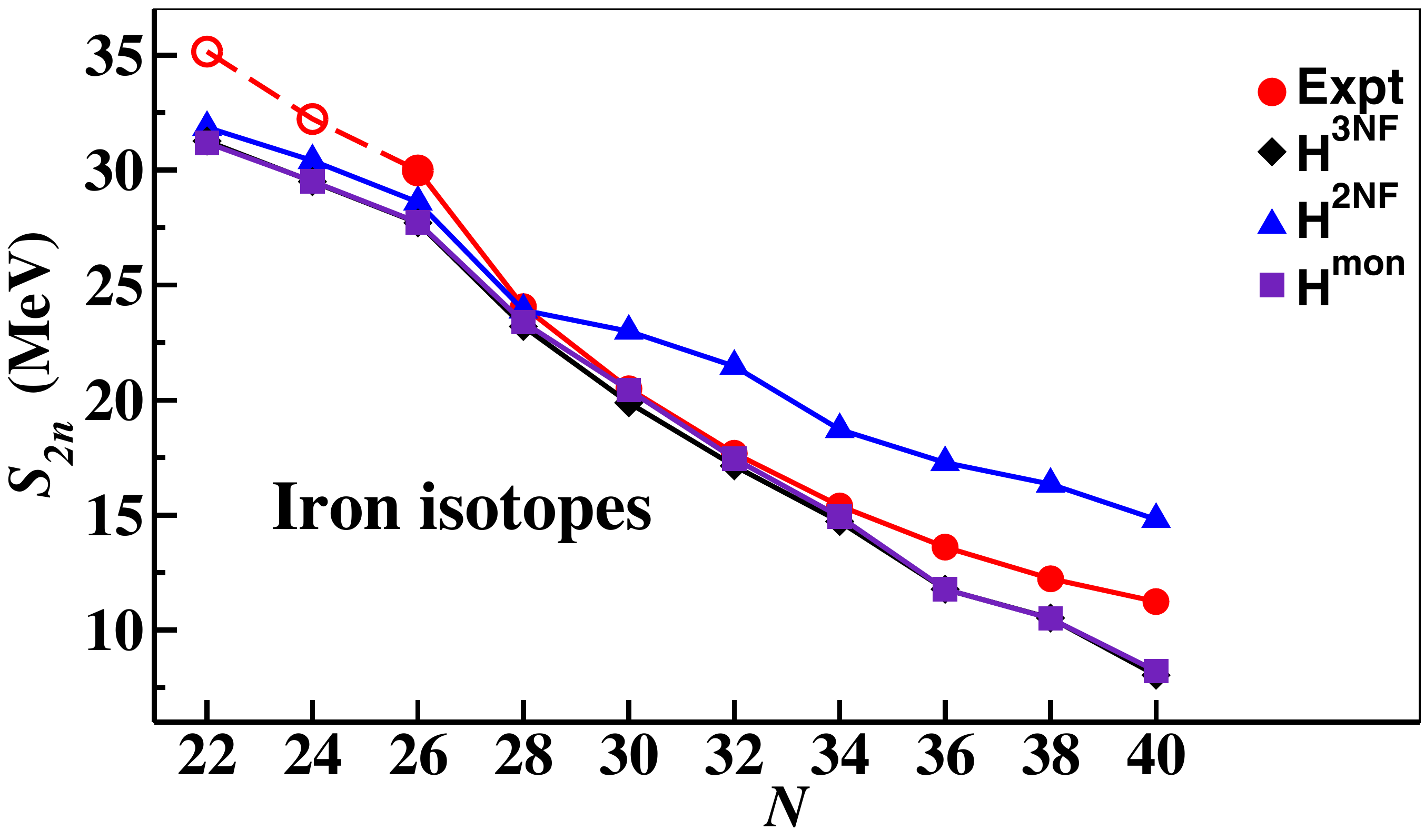}
\caption{Same as in Fig. \ref{S2nTi}, but for iron isotopes. See text
  for details.}
\label{S2nFe}
\end{center}
\end{figure}

Similar considerations follow from the inspection of
Figs. \ref{S2nCr},\ref{S2nFe}, where the experimental \cite{Audi03}
and calculated $S_{2n}$ for chromium and iron isotopes up to $N=40$
are shown, respectively.
We remind that empty red circles refer to estimated values reported
in Ref. \cite{Audi03}.

As can be seen, for these isotopes the observed $S_{2n}$ decrease from
$N=28$ to $N=30$ is no longer as steep as in calcium and titanium
isotopes, evidencing the quenching of the $N=28$ shell closure.

Once again the 3NF contribution, which is included in the monopole
component of $H^{\rm 3NF}_{\rm eff}$ and $H^{\rm mon}_{\rm eff}$,
provides a better reproduction of the experimental behavior at least
up to $N=34$. 

Finally, we examine the nickel isotopes whose study is pivotal to
understand the shell-closure properties of SM Hamiltonians.
As we have seen, the proton closure at $Z=28$ is eroded by the
increment of the number of valence neutrons approaching doubly-closed
$^{56}$Ni because of the collectivity induced by the proton-neutron
interaction.
Consequently, reproducing the evolution of  the spectroscopic
properties of nickel isotopes towards the shell closure may represent
a challenging test for the theoretical SP energies and TBMEs.

\begin{figure}[H]
\begin{center}
\includegraphics[scale=0.31,angle=0]{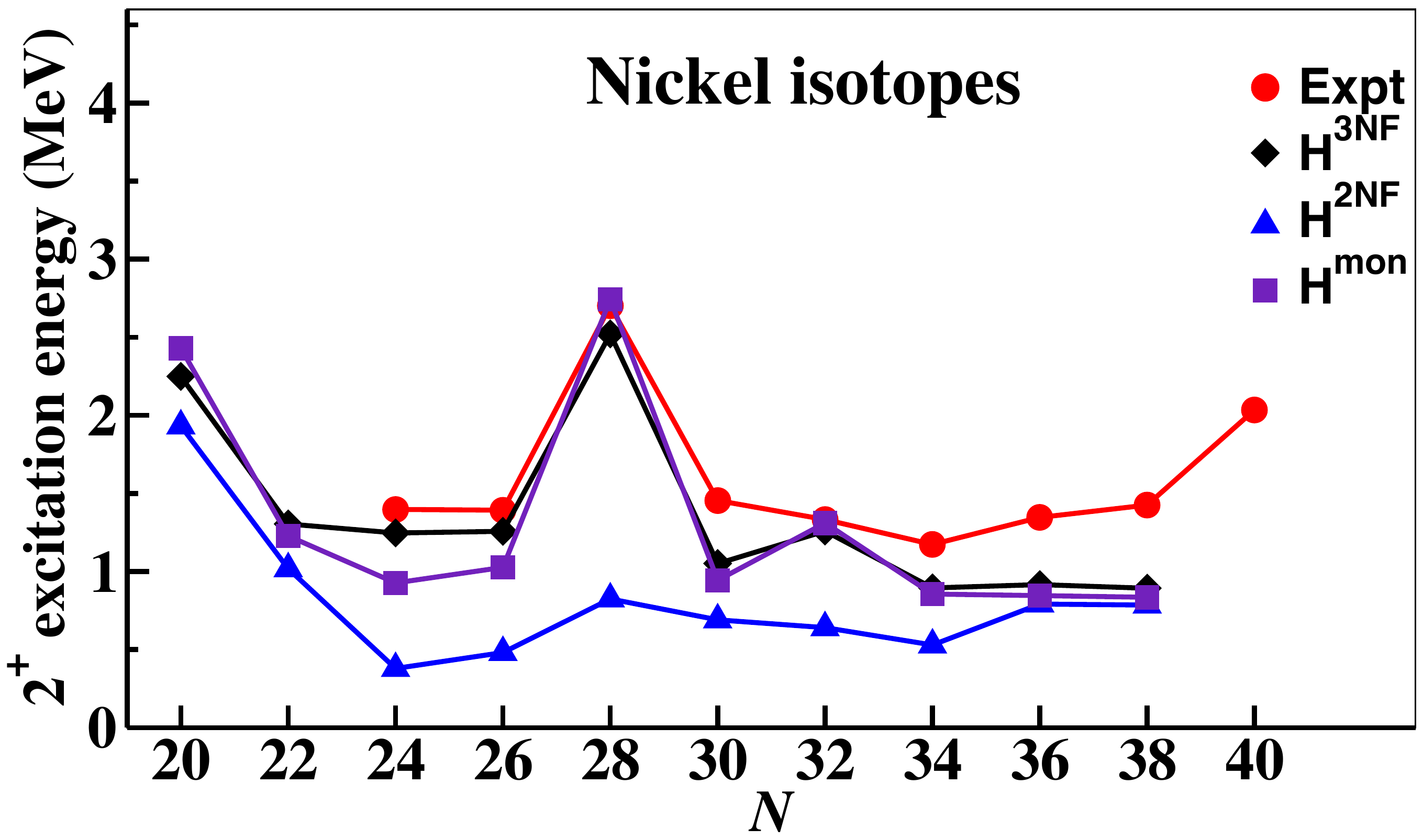}
\caption{ Same as in Fig. \ref{J2pTi}, but for nickel isotopes. See text
  for details.}
\label{J2pNi}
\end{center}
\end{figure}

In Fig. \ref{J2pNi} we show the behavior of the
  experimental $J^{\pi}=2_1^+$ excitation energies of nickel isotopes
  up to $N=40$ \cite{ensdf}, and the calculated ones up to $N=38$. 
  This different choice is due to the fact that the calculated values
  of the yrast $J^{\pi}=2^+$ excitation energies for $^{68}$Ni are
  larger than 7 and 5 MeV with and without 3NF contribution,
  respectively.
  Such an overestimated result overshoots the energy scale of
  Fig. \ref{J2pNi} - we have chosen to have the same scale in all
  similar figures for the sake of consistency - and is a mere
  consequence of the limitation of $fp$-shell model space to describe
  heavier systems.

As can be seen, the three effective Hamiltonians predict a shell
closure at $N=20$ ($^{48}$Ni), although less marked with $H^{\rm
  2NF}_{\rm eff}$, that confirms the ability of their monopole
components to provide a similar behavior in the identical-particle
channel.

Actually, both $H^{\rm 3NF}_{\rm eff}$ and $H^{\rm mon}_{\rm eff}$
results compare themselves quite well with $^{52,54,56}$Ni data, while
those obtained with $H^{\rm 2NF}_{\rm eff}$ exhibit a too strong
collective behavior, failing to reproduce the shell closure at
$N=Z=28$.
As a matter of fact, the comparison between the results obtained with
$H^{\rm 2NF}_{\rm eff}$ and $H^{\rm mon}_{\rm eff}$ evidences very
clearly that the correct shell evolution may be obtained only including 3NF
contributions in the monopole component of the SM Hamiltonian, the SP
energies being not sufficient to balance the collectivity induced by
the $T=0$ multipole component of the TBMEs.

\begin{figure}[H]
\begin{center}
\includegraphics[scale=0.31,angle=0]{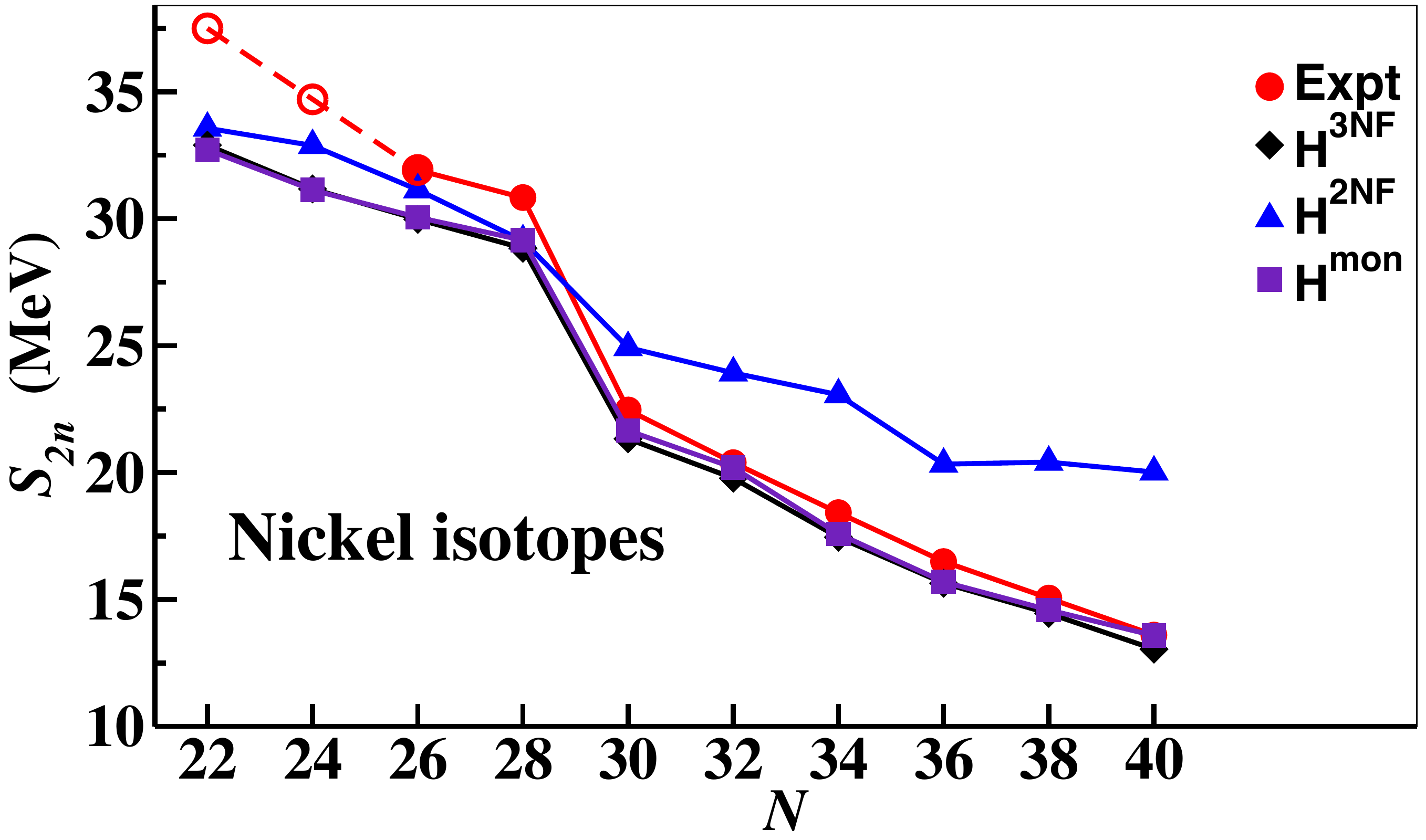}
\caption{Same as in Fig. \ref{S2nTi}, but for nickel isotopes. See text
  for details.}
\label{S2nNi}
\end{center}
\end{figure}

The same conclusions may be drawn from the inspection of the behavior
of the $S_{2n}$ as a function of the valence neutrons, which are reported
in Fig. \ref{S2nNi}.
For nickel isotopes the drop in energy between $N=28$ and $N=30$
appears again, and the experimental behavior \cite{Audi03} is
obtained correctly by means of $H^{\rm 3NF}_{\rm eff}$ and $H^{\rm
  mon}_{\rm eff}$.

\begin{figure}[H]
\begin{center}
\includegraphics[scale=0.35,angle=0]{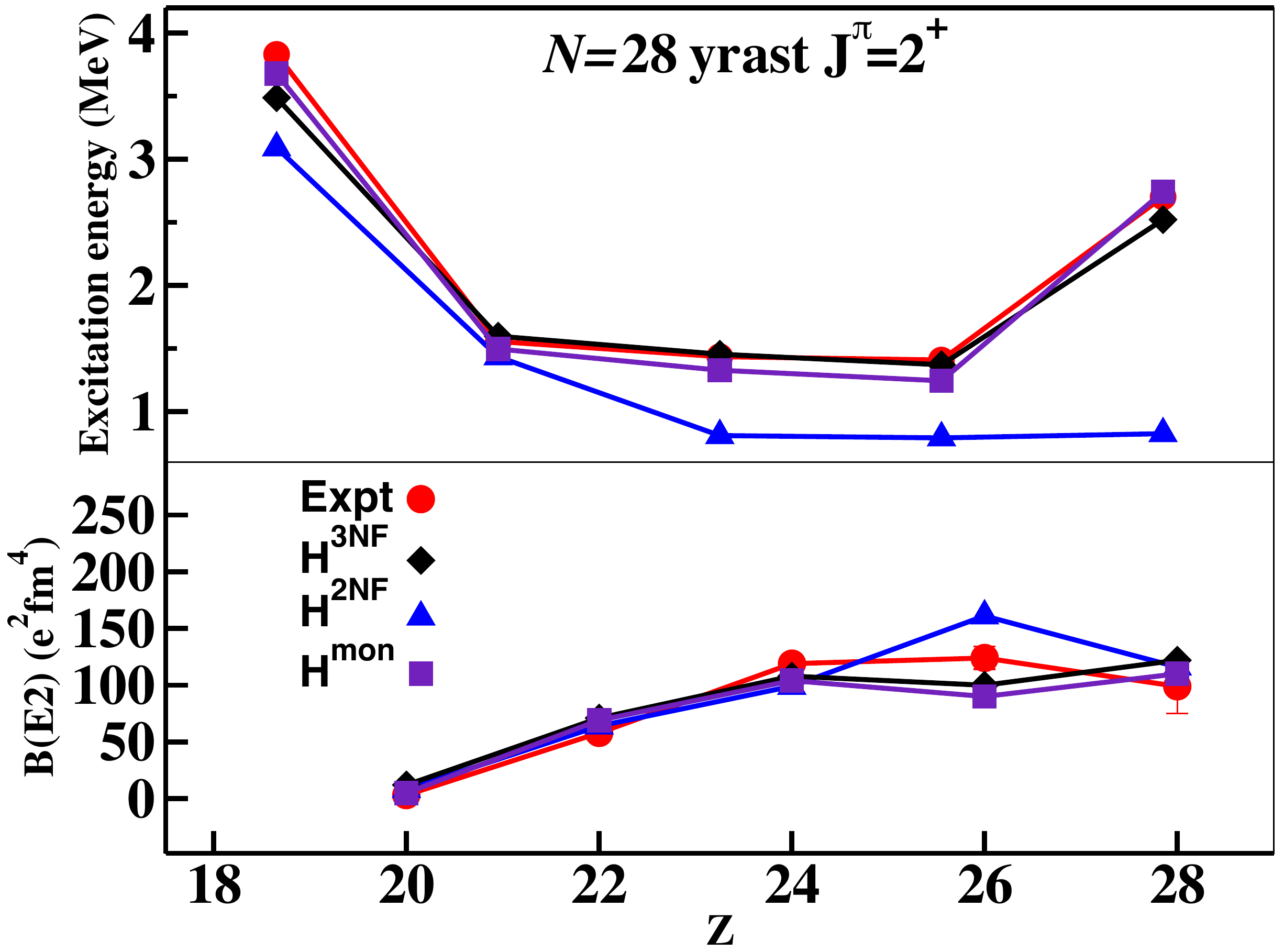}
\caption{Experimental and calculated excitation energies of the yrast
  $J^{\pi}=2^+$ states and $B(E2; 2^+_1 \rightarrow 0^+_1)$ transition
  rates for the $N=28$ isotones. See text for details.}
\label{N28}
\end{center}
\end{figure}

We conclude our discussion about the evolution of $N=28$ shell closure
summarising our results in Fig. \ref{N28}, where we have reported, for
the $N=28$ isotones, the experimental and calculated behavior of both
$J^{\pi}=2^+_1$ excitation energies and $B(E2; 2^+_1 \rightarrow
0^+_1)$ transition rates.
The proton and neutron effective charges to calculate the $B(E2)$s
have been obtained by way of many-body perturbation theory using only 2NF
vertices, and details of the derivation of effective SM one-body
operators can be found in Ref. \cite{Coraggio19a}.

As can be seen, the filling of the proton $0f_{7/2}$ orbital tunes the
collectivity at $N=28$ between the doubly closed $^{48}$Ca and
$^{56}$Ni, and the evolution of such a collective behavior is well
reproduced including 3NF contributions, but it is a failure by
considering only 2NF.

\section{Concluding remarks and outlook}
\label{conclusions}
In this paper we have presented the results of SM calculations for
$fp$-shell nuclei in the framework of the realistic shell model, starting
from chiral 2NF and 3NF, and deriving effective SM Hamiltonians within
the many-body perturbation theory.
These effective Hamiltonians account also, in their
  two-body matrix elements, of the different number of valence protons
  and neutrons characterizing each nucleus under investigation.

In particular, we have calculated the contribution at first order
in perturbation theory of a N$^2$LO  chiral 3NF potential to the
$H_{\rm eff}$, in order to study how it affects its monopole component
and the ability to describe the observed shell-closure properties of
$fp$ isotopic chains.
To this end, starting from two different class of $H_{\rm eff}$s - one
including 3NF contributions and the other one not - we have first
carried out an analysis of the effective single-particle energies for
calcium and nickel isotopes as a function of the valence-neutron
number.
This study has provided information about shell-closure properties
and  their dependence on the 3NF effects included in the monopole
components of $H_{\rm eff}$.

Successively, we have performed a full diagonalization of our $H_{\rm
  eff}$s for the calcium, titanium, chromium, iron, and nickel
isotopes, and focussed our attention on the shell evolution of the
excitation energies of the yrast $J^{\pi}=2^+$ states and the
two-neutron separation energies.

The conclusion of our study can be summarised as follows:
\begin{itemize}
\item Starting from realistic potentials, derived within the chiral
  perturbation theory, the role of the 3NF is fundamental to obtain SP
  energies and TBMEs that may reproduce the shell evolution as observed
  from the experiment.
\item The TBMEs of $H_{\rm eff}$ derived from 2NF only own deficient
  monopole components, which cannot balance the collectivity induced by
  higher multipole components in the proton-neutron channel. 
  The result is an erosion of the $N=28$ shell closure when the number
  of valence protons increases.
\item The central role of the monopole component of the $H_{\rm eff}$
  is testified by the fact that when it is subtracted from $H^{\rm 2NF}_{\rm
    eff}$, and substituted with the monopole of $H^{\rm 3NF}_{\rm
    eff}$, the observed shell evolution and the $N=28$ shell
  closure is restored.
\end{itemize}

The outlook of our future work points towards the improvement of the
derivation of $H^{\rm 3NF}_{\rm eff}$ by including higher-order
contributions with $3N$ vertices in the perturbative expansion of the
$\hat{Q}$ box, and the investigation of heavier systems in order to
assess the reliability of present approach in exotic neutron-rich
nuclear systems.

\section*{Acknowledgements}
This work has been supported by he National Key R\&D Program of China
under Grant No. 2018YFA0404401, the National Natural Science
Foundation of China under Grants No. 11835001, No. 11320101004, and
No. 11575007 and the CUSTIPEN (China-US Theory Institute for Physics
with Exotic Nuclei) funded by the US Department of Energy, Office of
science under Grant No. DE-SC0009971. We acknowledge the MARCONI
Supercomputer at CINECA, under the CINECA-INFN agreement, and
the High-performance Computing Platform of Peking University for
providing computational resources.

\bibliographystyle{apsrev}
\bibliography{biblio.bib}

\end{document}